\documentclass{aa}
\usepackage{natbib}
\bibpunct{(}{)}{;}{a}{}{,} 
\input{bib.def}

\usepackage{xcolor,colortbl}
\definecolor{MyRed}{rgb}{0.9,0.0,0.0} 
\definecolor{MyPink}{rgb}{0.8,0.3,0.5} 
\definecolor{MyMediumBlue}{rgb}{0.7,0.72,1.0} 
\definecolor{Sven}{rgb}{0.0,0.4,0.4} 
\definecolor{Grey}{rgb}{0.5,0.5,0.5} 
\definecolor{MyGreen}{rgb}{0.0,0.9,0.5} 
\definecolor{Henrik}{rgb}{0.2,0.4,0.95}
\definecolor{HenrikII}{rgb}{0.8,0.8,0.1}
\definecolor{HenrikIII}{rgb}{0.5,0.9,0.7}
\definecolor{Shahin}{rgb}{1.0,0.11,0.0}

\newcommand{\swedt}[1]{{{#1}}}

\newcommand{\hea}[1]{{#1}}

\newcommand{\sjedt}[1]{{#1}}

\usepackage{graphicx}
\usepackage{txfonts}

\usepackage{ulem}
\begin{document} 

   \title{The Sun at millimeter wavelengths}
   \subtitle{II. Small-scale dynamic events in ALMA Band 3}   

   \author{Henrik Eklund* \inst{1,2}
          \and
           Sven Wedemeyer \inst{1,2}
          \and
           Mikolaj Szydlarski  \inst{1,2}
          \and
           Shahin Jafarzadeh  \inst{1,2}
          \and 
           Juan Camilo Guevara G\'omez \inst{1,2}
  }
   
  \authorrunning{Eklund {et~al.}}
   \institute{Rosseland Centre for Solar  Physics, University of Oslo, Postboks 1029 Blindern, N-0315 Oslo, Norway
            \and
            Institute of  Theoretical Astrophysics, University of Oslo, Postboks 1029 Blindern, N-0315 Oslo, Norway \\
            \email{henrik.eklund@astro.uio.no}
} 
 
\def\corrAuthor{Henrik Eklund}

   \date{Received --- ; accepted --- }

\abstract{Solar observations with the Atacama Large Millimeter/sub-millimeter Array (ALMA) facilitate studying the atmosphere of the Sun at chromospheric heights at high spatial and temporal resolution at millimeter wavelengths.}
{ALMA intensity data at mm-wavelengths are used for a first detailed systematic assessment of the occurrence and properties of small-scale dynamical features in the quiet Sun.}
{ALMA Band~3 data ($\sim 3$\,mm / $100$\,GHz) with spatial resolution $\sim 1.4$ -- $2.1$ arcsec and a duration of $\sim 40$\,min are analysed together with SDO/HMI magnetograms.
The temporal evolution of the mm-maps is studied to detect pronounced dynamical features which are connected to dynamical events via a k-means clustering algorithm. The physical properties of the resulting events are studied and it is explored if they show  properties consistent with propagating shock waves. For this purpose, observable shock wave signatures at mm wavelengths 
are calculated from one- \hea{and three-}dimensional model atmospheres.}
{There are 552 dynamical events detected with an excess in brightness temperature ($\Delta T_\text{b}$) of at least $\geq 400$\,K. The events show a large variety in size up to $\sim9\arcsec$, amplitude $\Delta T_\text{b}$ up to $\sim1200$\,K with typical values between $\sim450$ -- $750$\,K and lifetime at FWHM of $\Delta T_\text{b}$ between $\sim43$ -- $360$\,s, with typical values between $\sim 55$ -- $125$\,s. 
Furthermore, many of the events show signature properties that suggest that they are likely produced by propagating shock waves.}
{There are a lot of small-scale dynamic structures detected in the Band~3 data, even though the spatial resolution sets limitations of the size of events that can be detected. The amount of dynamic signatures in the ALMA mm data is very low in areas with photospheric footpoints with stronger magnetic field\hea{s}, which is consistent with the expectation for propagating shock waves.}

   \keywords{Sun: chromosphere -- Sun: radio radiation -- Sun: atmosphere -- Sun: magnetic fields -- shock waves -- techniques: interferometric}

   \maketitle

\section{Introduction}
\label{sec:introduction}

The Atacama Large Millimeter/sub-millimeter Array (ALMA) provides new \swedt{diagnostic possibilities for studying the dynamical nature} of the solar chromosphere \sjedt{at millimeter wavelengths} 
\swedt{due to} 
\swedt{its high spatial and temporal resolution}.
The mm-wavelength radiation observable with ALMA origins at chromospheric heights \citep[][and references therein]{2016SSRv..200....1W}  from free-free continua emission under the conditions of local thermodynamic equilibrium (LTE). The measured continuum intensities (and equivalently the continuum brightness temperatures) is therefore expected to depend linearly on the local thermal gas temperature, or to be more precise, the electron temperature.

\sjedt{The solar chromosphere is a highly dynamic and structured region of the Sun's atmosphere} 
\swedt{that} 
features a variety of phenomena on different spatial scales.
\sjedt{
In particular, small-scale  structures have been shown to evolve often at short time scales, on the order of a few seconds or less \citep{2011ApJ...736L..24O,2012A&A...543A...6M,2017ApJS..229....7G,2017ApJS..229....9J,2017ApJS..229...10J}. 
Thus, both high spatial and temporal resolution are needed to fully understand this dynamic region. It is also worth nothing that most of the commonly employed chromospheric diagnostics (at ultraviolet, optical, and infrared wavelengths) are formed under non-LTE conditions \citep{2017SSRv..210..109D,2019ARA&A..57..189C}, which as a result, renders the meaningful interpretation of such observations challenging. Thus, by promising to provide  direct measurements of temperatures under LTE conditions at high temporal resolution, ALMA is an excellent tool for studying dynamic small-scale structures in the solar chromosphere \citep[see, e.g.,][]{2020arXiv200804179G}.}\\ 
Observations of the quiet-Sun internetwork chromosphere in the \ion{Ca}{II}\,K line revealed 
a bright mesh-like pattern of elongated structures with fainter intermediate areas outside strong magnetic field concentrations \citep{2006A&A...459L...9W}.  
The typical mesh size of the observed pattern, which is produced by the interaction of propagating shock waves,  
was determined to be on the order of $1\arcsec.95$.
Numerical simulations of the solar atmosphere predict that a corresponding pattern should be observable with ALMA \citep[][and references therein]{2007A&A...471..977W,2015A&A...575A..15L,2016SSRv..200....1W}.

The corresponding shock wave signatures at mm wavelenghts as observable with ALMA have been predicted by \citet{2004A&A...419..747L, 2006A&A...456..713L} using one-dimensional models \hea{and by \cite{2020arXiv200805324E} using a three-dimensional model \citep[cf.][]{2007A&A...471..977W}.} 
They found large brightness temperature variations with amplitudes of thousands of Kelvin. 
Studies of observational ALMA data, on the other hand, indicate smaller temperature variations on the order of several hundred Kelvin \citep[see, e.g.][]{2020A&A...634A..86P,2018A&A...619L...6N, 2017A&A...605A..78A, 2020A&A...635A..71W, 2019ApJ...877L..26L}.

We note that the spatial resolution of available ALMA band~3 data used in publications is around $2\,\arcsec$ or below \citep[e.g.,][]{2019ApJ...877L..26L}  with the minor beam axis down to $1.4\,\arcsec$ 
\citep[see][and references therein]{2020A&A...635A..71W}
and is thus on the same order as the aforementioned typical spatial scale of the shock wave induced mesh-like pattern. The visibility of this pattern and with it the amplitude of observed brightness temperature variations therefore critically depend on the angular resolution achieved with ALMA. At the same time, ALMA provides a very useful tool to study and understand shock waves and the general dynamics of the chromosphere given that the measured brightness temperatures are expected to be closely related to the local electron temperature in the probed layer \citep[][and references therein]{2016SSRv..200....1W}. 
Accurate statistics of the spatial and temporal scales as well as the amplitudes of the brightness temperature variations can also be used as a feedback to numerical models of the solar atmosphere with the aim to refine and make the simulations even more accurate and realistic. 

Here, we analyse one of the first regular Band~3 \hea{($\sim3$\,mm / $100$\,GHz)} observations of the Sun with ALMA (Cycle\,4), which has been described in \citet{2020A&A...635A..71W}. 
\hea{This high cadence ($2$\,s) time series}
\swedt{is used for a statistical study of detected brightness temperatures signatures of small-scale dynamic events.}  
\swedt{An important question in this regard is if the detected signatures are caused by propagating shock waves in the solar chromosphere.} 
This paper is structured in the following way: In Sect.\,\ref{sec:methods}, the properties and the processing of the observational data is described. In Sect.\,\ref{sec:results}, the statistics of the detected dynamical features are presented. In Sect.\,\ref{sec:disc} we further discuss the results \hea{and, with the support of numerical simulations, investigate the possibility that the  detected events are signatures of propagating shock waves.}
\swedt{In that respect, the limitations of the observational data are addressed.}  
In Sect.\,\ref{sec:conc}, we summarize and conclude the results and give a brief outlook on future work.

\section{Methods}\label{sec:methods}

\subsection{Observational data}\label{sec:methods_obs_data}
The observational data studied in this paper was obtained with ALMA in receiver band~3 on December 22nd, 2016 (project ID: 2016.1.00423.S), targeting a quiet Sun region \hea{with a few magnetic network elements} close to solar disc centre. The observations took place between $14$:$19 \text{UT} - 15$:$07 \text{UT}$ parted in four time \hea{scans in blocks} of $10$\,min each, separated by calibration gaps of $\sim2.5$\,min. The data \hea{set} is processed using the Solar Alma Pipeline (SoAP). The specifics of SoAP will be explained in detail in a forthcoming paper \citep[Szydlarski~et~al. in prep., see also][]{2020A&A...635A..71W}. In summary, 
\swedt{data set is calibrated (with a script }
delivered together with the data from the ALMA Science Archive) followed by deconvolution using the Multiscale CLEAN algorithm \citep{2008ISTSP...2..793C} and primary beam correction, which results in   interferometric intensity maps. These are combined (`feathered')  with single dish total power measurements. The intensities are transformed into brightness temperatures under the assumption of the Rayleigh-Jeans law and formation in local thermodynamic equilibrium \citep[see, e.g.,][]{2013tra..book.....W} \hea{through the relation} 
\hea{\begin{equation}\label{eq:tb}
    T_\text{b}=\frac{c^2 I_\nu}{2k_\text{B} \nu^2}
\end{equation}}
\hea{where $c$ is the speed of light and $k_\text{B}$ is the Boltzmann constant.} The images are then combined into a time series. The final dataset has a cadence of 2\,s, with a total of 1200 time steps. For an in depth analysis and details on this specific dataset, see \citet{2020A&A...635A..71W}, where the quality and limitations of the data is evaluated. \sjedt{The same dataset has also been analysed in \citet{Jafarzadeh2020}, where global oscillations in the time-series of images (along with other ALMA datasets) have been studied.}

\begin{table}[b!]
\caption{Span of wavelength and frequency of the sub-bands of ALMA spectral band~3.}
\label{tab:frequencies}
\centering
\begin{tabular}{ccccccc}
\hline
\multicolumn{7}{c}{Band 3}\\
\hline
&\multicolumn{3}{c}{Wavelength [mm]}&\multicolumn{3}{c}{Frequency [GHz]}\\
Sub-band&min&mid&max&min&mid&max\\
\hline                
SB1 & 3.189&3.224& 3.259& 92.0 &93.0&  94.0  \\ 
SB2 & 3.123&3.156& 3.189& 94.0 &95.0&  96.0  \\
SB3 & 2.828&2.856& 2.883&104.0 &105.0& 106.0  \\
SB4 & 2.776&2.802& 2.828&106.0 &107.0& 108.0  \\
\hline
\end{tabular}
\label{tab:almasubbands}
\end{table}

For solar observations, ALMA receiver band~3 is currently set up with four spectral sub-bands as indicated in Table \ref{tab:frequencies}.
\hea{The formation height of the mm-wavelength radiation is wavelength dependent.
ALMA band~3 is thought to be formed in the upper chromosphere \citep[][and references therein]{2020A&A...635A..71W, 2017SoPh..292...88W}. 
The sub-bands are grouped pairwise where SB1 is adjacent to SB2 and SB3 is adjacent to SB4. Differences between the sub-bands can in principle be used to study the small scale structure} along the line of sight 
\citep[see e.g.,][]{2019A&A...622A.150J,2019ApJ...875..163R}
although simulations imply that the differences are typically small \citep[see, e.g.,][]{2020arXiv200805324E}. 
In this work, however, we choose to combine all spectral channels of all four sub-bands in the deconvolution process when producing the intensity maps as this results in better sampling of the Fourier space and thus a higher signal-to-noise ratio, 
\swedt{which is advantageous for the detection and analysis of small-scale dynamic features}.

The output of the image reconstruction is cut off at a radius of $32.8\arcsec$ from the centre field of view (FOV). The cut-off point corresponds to 0.3 times the main beam power response. The pixel size of the final images is $0.32\arcsec$. 
\hea{The ``clean'' beam, i.e. a Gaussian fit to the main lobe of the point spread function (PSF) of the interferometric array constitutes the resolution element.} The Full Width Half Maximum (FWHM) of the clean beam varies slightly by time but has a mean value of about $1.4\arcsec$ and $2.1\arcsec$ along the minor and major axis respectively, with a beam position angle of $68^\circ$. 

Magnetic field measurements are taken with the Helioseismic Magnetic Imager (HMI; \citealt{2012SoPh..275..229S}) on board the Solar Dynamics Observatory (SDO; \citealt{2012SoPh..275....3P}) . The magnetograms\hea{, measuring the Fe I line (6173 Å) at photospheric heights,} have a cadence of $36$ seconds and are co-aligned with the ALMA observations. The 
\swedt{SDO images are re-sampled to a pixel size of}  
$0.32"$ to match the ALMA data. \sjedt{For the spatial co-alignment, the ALMA images were cross-correlated with a combination of the $170$\,nm and $30.4$\,nm channels of the  SDO's Atmospheric Imaging  Assembly (AIA; \citealt{2012SoPh..275...17L}).} 

\subsection{Event definition and selection}
\label{sec:temp_var_obs_data}

\begin{figure}[tp]
\includegraphics[width=\columnwidth]{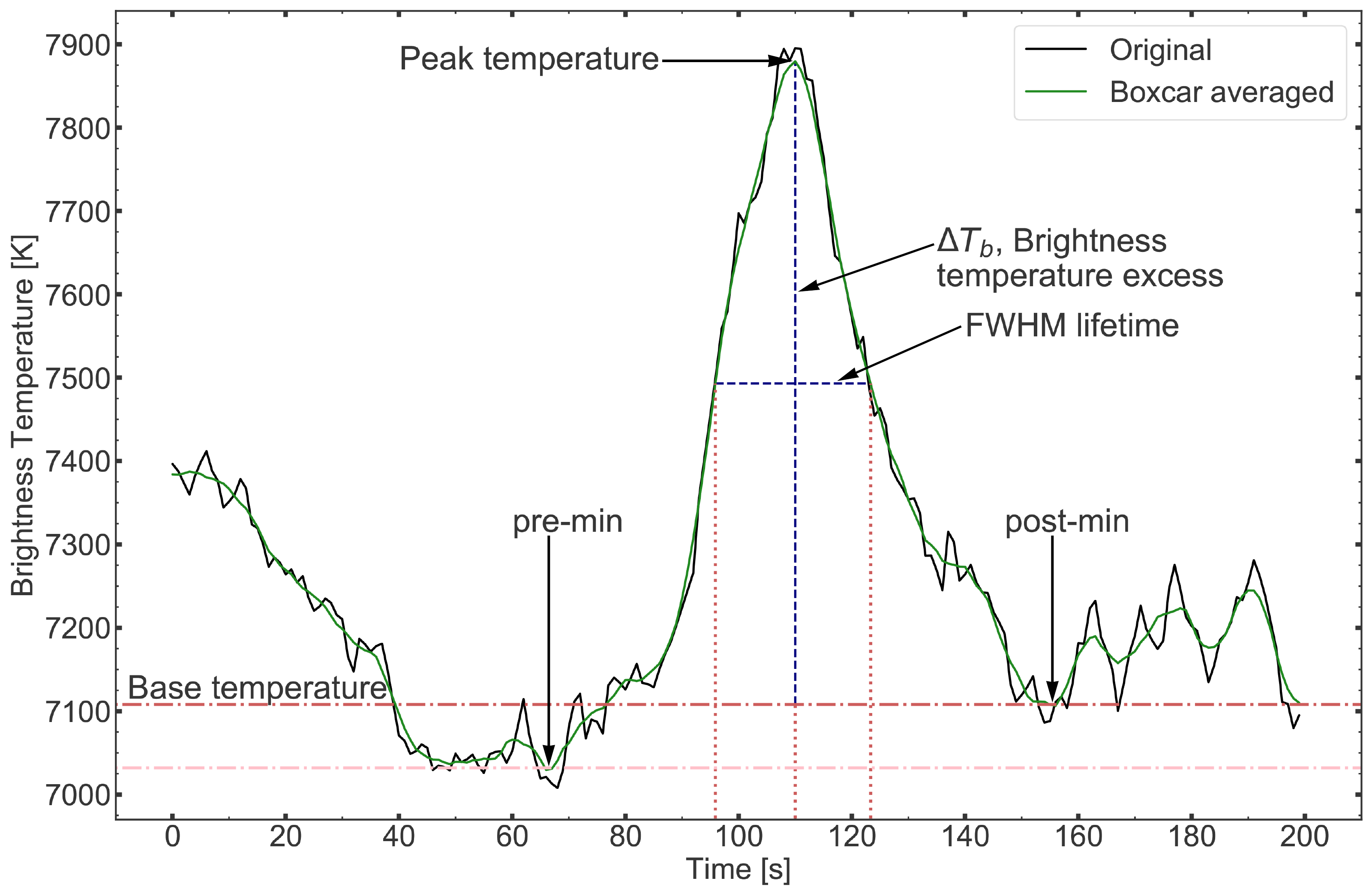}
\caption{Definition and selection criteria for a dynamic event. Temporal evolution of the brightness temperature at a selected pixel exhibits a peak enclosed by two minima (pre-min. and post-min.). The green graph shows the boxcar average of the original data (black graph). The blue dashed lines marks the peak amplitude and FWHM lifetime. The horizontal red dashed-dotted line marks the temperature of the local minima with highest temperature, in this case the post-peak minima, which is the base brightness temperature that is considered for the specific event. The pink dashed-dotted line shows the temperature of the other minima, in this case the pre-peak minima. The red dotted lines mark the times where the temperature is at FWHM and peak.}
\label{fig:shockillustration}
\end{figure}

The observational data are searched for signatures of small-scale dynamic events. 
For that purpose, the brightness temperature time evolution ($T_\text{b} (t)$) of each pixel is studied. To remove high frequent noise, the data is box-car averaged \hea{in time} with a size of $7$ time steps $(14\,s)$. 
The resulting smoothed data are then searched for peaks in brightness temperature
\hea{at every pixel composed of an increase followed by a decrease in brightness temperature (cf. Fig.~\ref{fig:shockillustration}) with an amplitude of at least $400$\,K}.  
The amplitude of the peak in brightness temperature excess is \hea{measured} as the difference to the base temperature level as determined by the closest significant local minima of $T_\text{b} (t)$ at the given spatial position. To be on the conservative side, the minimum with the smaller $T_\mathrm{b}$ difference to the peak is chosen. The lifetime of an event is then determined as the FWHM of the brightness temperature excess peak (see Fig.~\ref{fig:shockillustration}).

The pixels showing such a temporary rise in brightness temperature are grouped together using the k-means clustering method \citep{Everitt:1972aa}. 
Each resulting cluster represents an individual event. The k-means clustering method parts multidimensional data into a predefined amount of clusters and then assigns the pixels to clusters depending on the euclidean distance in spatial and temporal coordinates with respect to the mean value for each cluster. The suitable amount of clusters is determined by a combination of the following two methods:

($i$) Start with a small number of clusters and perform the k-means clustering. Calculate the entropy (i.e., the mean of the spreading within the clusters). Iteratively repeat the process with an increasing number of clusters. At last pinpoint the number of clusters where the entropy does not drop significantly compared to previous steps with smaller number of clusters. This method is commonly referred to as the "elbow method". 

($ii$) Comparing each data point to its own cluster and to the closest neighboring cluster to see which cluster it is most closely related to. This method is commonly referred to as the "silhouette method".

In the central parts of the FOV, the events are more easily separated than when going towards the outer parts of the FOV. We conclude that this effect is caused by the radial increase of $T_\mathrm{b}$ uncertainties resulting from the image reconstruction process of the interferometric measurements. The uncertainties becomes very pronounced in the outer parts of the FOV. Because of the lower reliability of the data in these parts, the k-means clustering is only applied to the innermost parts within a radius of $15\arcsec$.

In that inner region, all events are taken into account that ($i$)~show clear peaks with a brightness temperature excess of more than $400$\,K and ($ii$)~are fully represented in a single time scan. Thus, the $10$ minutes length of the time scans restricts the amount of events detected. Since the imaging procedures of solar ALMA data are still in an early stage, the criteria are defined more conservatively for this initial study in order to minimize the number of false detection.

\section{Results} 
\label{sec:results}

\subsection{\hea{Events with peaks in the temporal evolution of the brightness temperature}}\label{sec:results:peak_clustering}

\begin{figure*}[tp!]
\includegraphics[width=\textwidth]{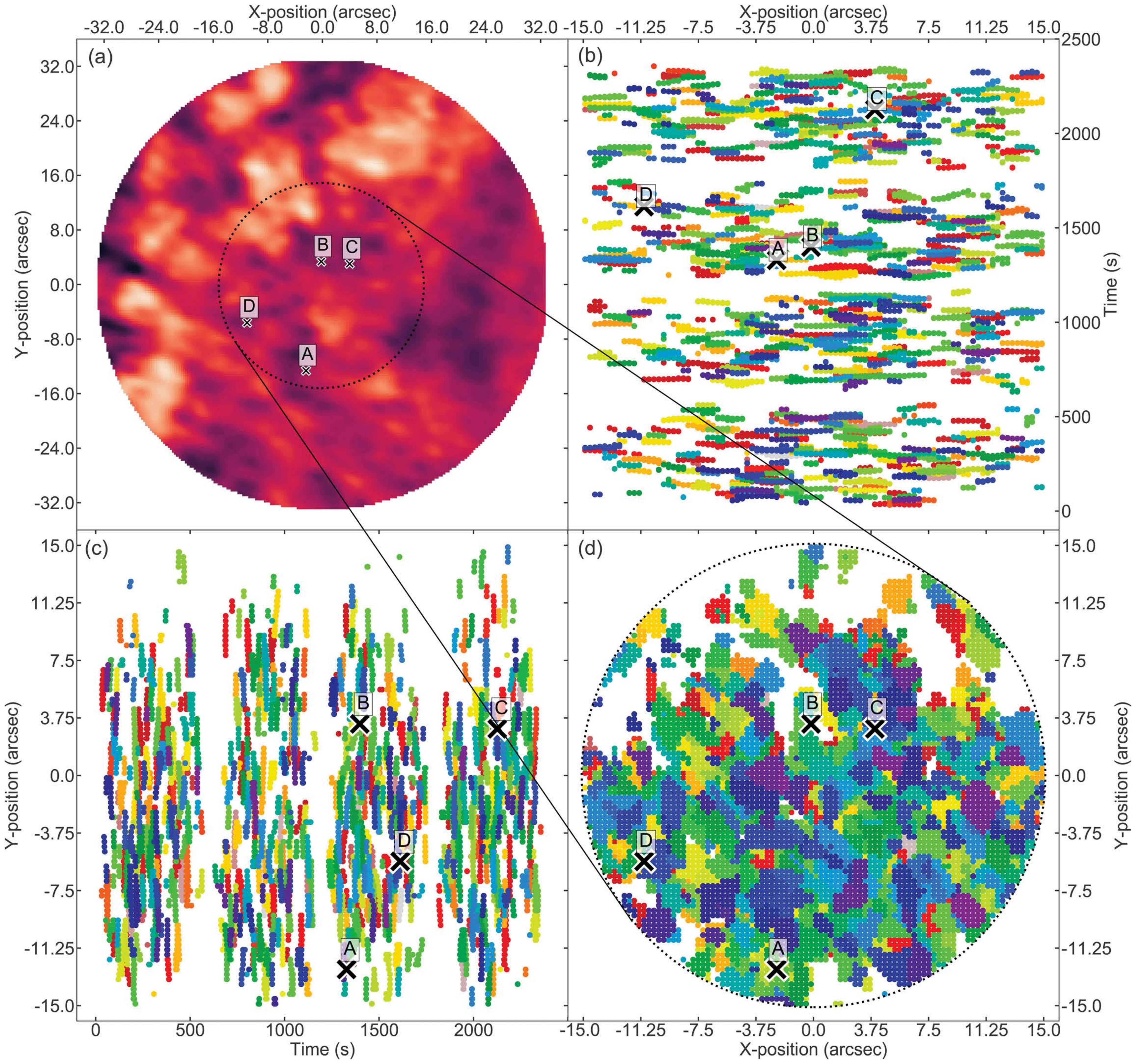}
\caption{
\hea{Spatial and temporal position of all 552 detected temperature peak events within 15 arcseconds from the centre of the field of view and with a magnitude of at least 400\,K. 
\textbf{a)} Brightness temperature map of the first frame of the dataset (2s integration). The black dotted circle marks a radial distance of 15 \arcsec from centre of the field of view. 
\textbf{b)}~The x-axis position and 
\textbf{c)}~the y-axis position
of all brightness temperature peaks plotted against time, respectively.
The events are all colored individually, showing the same color for the same event in all panels.  \textbf{d)} The x-axis and y-axis positions of all temperature peaks. The locations of a few selected events (A,B,C and D) that are studied in detail in Sect.~\ref{sec:event_detailed_study} are marked with crosses in all panels.}}
\label{fig:events_coordinates}
\end{figure*}

A total number of $552$ events  are found within the inner parts of the FOV with a radius of $15$\arcsec (see Sect.~\ref{sec:temp_var_obs_data} for the selection criteria). 
The individual events are well distributed over almost the entire selected part of the field of view \hea{and all four time scans} as can be seen from the spatial and temporal coordinates of the individual events in Fig.~\ref{fig:events_coordinates}.

There are some parts, dominated by the top part as well as smaller segment in the bottom of the inner region that do not show any events with brightness temperature variations more than $400$\,K \hea{(Fig.~\ref{fig:events_coordinates}d).} This is because of the presence of strong magnetic fields and is studied further in Sect.~\ref{sec:magnetic field}.

\begin{figure}[tp!]
\includegraphics[width=\columnwidth]{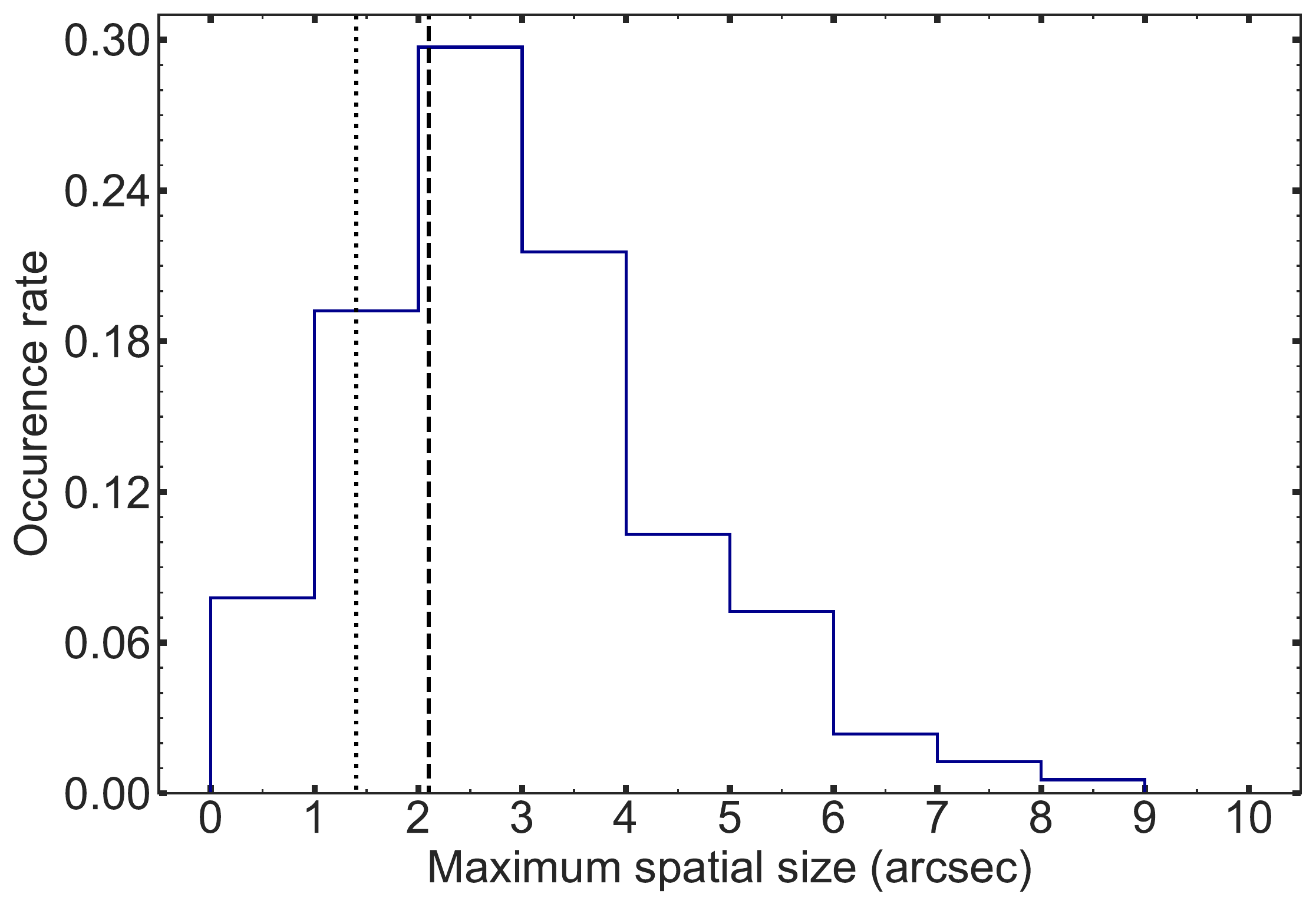}
\caption{Spatial size distribution of the 552 detected events, represented in bins with width of\,$1.0\arcsec$.
The vertical dotted and dashed lines mark the sizes of the clean beam minor and major axis, respectively.}
\label{fig:size_histogram}
\end{figure}

\hea{The events have spatial sizes up to about $9\,\arcsec$ which corresponds to $\sim 4.3$ -- $6.4$ times the clean beam. The spatial size of an event is indicated by the maximum euclidean distance between the pixels within the event. 
The majority of events are however smaller than $\sim2.5\arcsec$, which can be seen in Fig.\,\ref{fig:size_histogram} where the distribution of spatial sizes is presented.} From the peak value \hea{at $2.5\arcsec$,} there is a steady decrease in number of events towards larger sizes. Events smaller than the clean beam axes can not be spatially resolved, resulting in a decrease in the occurrence rate towards smaller sizes. 
Pronounced events might be detected despite being not spatially resolved but a larger number of events with smaller sizes is expected.

For a specific event, each pixel has a well defined time for the maximum peak brightness temperature which is plotted in panel b and c in Fig.\,\ref{fig:events_coordinates}.
The lifetime and the magnitude of the peak brightness temperature ($\Delta T_\text{b}$) varies between the pixels within a single event. This variation is caused by a combination of the inherent properties of the evolution of different parts of the events and the limited spatial resolution, which effectively smooths out the signatures. See a more detailed discussion of this issues in Sect.~\ref{sec:disc}. 
For this study, the overall lifetime of the event is determined as FWHM of the peak within the event with the largest $T_\mathrm{b}$ excess (cf. Fig.~\ref{fig:shockillustration}).  
The resulting values of lifetime and excess in $T_\mathrm{b}$ vary largely among the individual events. 
The lifetimes ranging between $\sim43$ -- $360$\,s and the brightness temperature excess $\Delta T_\text{b}$ between the cut-off limit of $400$\,K up to $\sim1200$\,K. 
A statistical analysis of these results is presented in Sect.~\ref{sec:Statistical study of events}.

\subsection{Magnetic field strength and network/internetwork mask \hea{of the entire FOV}}
\label{sec:magnetic field}
\begin{figure}[tp!]
\centering
\includegraphics[width=8.8cm]{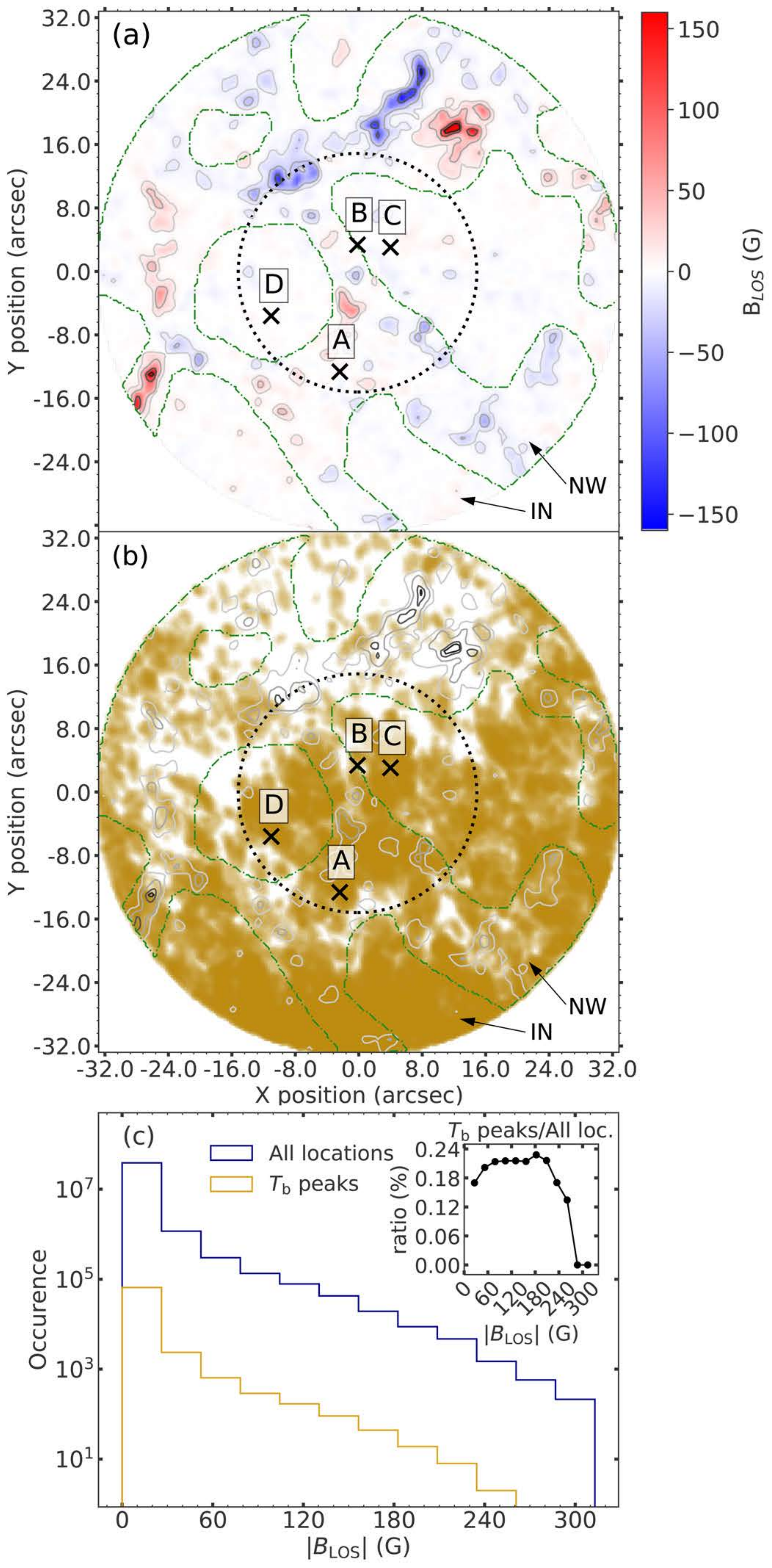}
\vspace*{-2mm}
\caption{Spatial comparison between brightness temperature dynamics in ALMA and SDO/HMI magnetogram. \textbf{a)} The time averaged SDO/HMI magnetogram for the whole ALMA FOV is presented as blue-white-red shades with the value range saturated at $[-160, 160]$\,G. The absolute line-of-sight (LOS) magnetic field strength is outlined as grey contour lines with the levels $10$, $30$, $80$ and $120$\,G. The green dotted-dashed lines marks the mask defining the borders between network and internetwork regions. The dotted ring marks the radial distance of $15\arcsec$ from the centre of the FOV. The spatial positions of the selected events A$-$D (Sect.~\ref{sec:event_detailed_study} $\&$ Appendix~\ref{sec:appendix:events}) are marked for reference. \textbf{b)}~The spatial position of all instances in the entire FOV with a $T_\text{b}$ excess of more than 400\,K and a lifetime shorter than $200$\,s, at any time in the data, are marked with orange-brown dots. The white spaces are thus locations without any events and a single pixel can show several $T_\mathrm{b}$ instances. 
The gray contour outlines the SDO/HMI magnetogram, whereas the green dotted-dashed contours, the dotted ring and the crosses A$-$D marks the same as in panel~a. 
\textbf{c)}~Histograms of the absolute LOS magnetic field strength at the locations and times of the $T_\mathrm{b}$ peaks marked in panel~b (orange-brown) and for the entire HMI data set (blue). The inset plot shows the ratio of the two histograms.}
\label{fig:mag_cont_mask}
\end{figure}

The observed quiet Sun region is composed of magnetic network (NW) and interwork (IN) regions \citep{2020A&A...635A..71W}. 
The  magnetic field strength of the entire FOV, averaged in time over the ALMA observation period, is shown in the SDO/HMI magnetogram in Fig.~\ref{fig:mag_cont_mask}a and ranges from -160\, to 230\,G. 
The strongest magnetic fields are present in the top parts of the FOV as well in the lower left corner. In the top right, there are two nearby locations with concentrations of opposite polarity. Between these locations, there are magnetic loop structures observable in many of the SDO-AIA channels. This group of compact loops is also seen in the ALMA Band~3 maps \citep{2020A&A...635A..71W}. 
The network/inter-network mask by \cite{2020A&A...635A..71W} is overplotted in Fig.\,\ref{fig:mag_cont_mask}. 
\swedt{It was composed based on} 
time-averaged observational maps from SDO-AIA in $160$ and $170$\,nm and saturated SDO/HMI magnetograms in combination with the ALMA maps. The locations of the stronger magnetic fields and their polarities are roughly \hea{stationary} during the time period of the ALMA observations and therefore an average over the whole observational period is used when defining the different areas.

\swedt{Using a  network/inter-network mask  is a commonly  used method for distinguishing between different types of regions, but
as such represents a compromise across the considered observation period and different imprints of the network in the considered atmospheric layers. 
Consequently, such a mask does not always adequately represent the  conditions at the exact time and location of a specific event  like for the  dynamical events discussed here. The mask is shown nonetheless in order to connect to the previous paper by \citet{2020A&A...635A..71W} and in order to facilitate comparisons with other studies using network/internetwork masks 
\citep[e.g.][]{2009A&A...497..273L}.
} 

In Fig.~\ref{fig:mag_cont_mask}b, all pixels that show a peak with brightness temperature excess with a minimum amplitude of $400$\,K and a lifetime shorter than $200$\,s \swedt{in the entire FOV and in all} of the four time scans are marked. The contours of the time-averaged magnetic field strength and the network/internetwork mask are overplotted for direct comparison.
The occurrence of $T_\mathrm{b}$ peak signatures is much lower in the upper and left parts of the FOV, where the magnetic field strength is on average stronger, as well in a few patches at a radial distance $\sim15\arcsec$ towards the bottom. 
It should be emphasised that the magnetic field structure varies significantly in time on small spatial scales with non-averaged \hea{line-of-sight} (LOS) magnetic field strength spanning between $[-260, 300]$\,G. 
As a result, some $T_\mathrm{b}$ peaks appear to occur at locations with stronger time-averaged LOS magnetic field strength (Fig.\,\ref{fig:mag_cont_mask}b), while the actual 
magnetic field strength at the exact time (and before) the occurrence of the peak is typically much lower.  
For this reason, a histogram for the LOS magnetic field strength at the exact location and the  time of each $T_\mathrm{b}$ peak is shown in Fig.\,\ref{fig:mag_cont_mask}c in comparison to the distribution of all magnetic field strength values in the data set.  
The ratio of the two distributions seems to remain more or less constant  for  LOS magnetic field strengths below $\sim180$\,G  but decreases with increasing magnetic field strengths beyond that point. This finding implies that the occurrence of brightness temperature peaks is reduced at locations with a LOS magnetic field strength of $\gtrsim180$\,G but the number of detections, on which this conclusion is based, is rather low. Furthermore, while it is uncertain to what degree one can relate the SDO/HMI magnetograms measured at photospheric levels to the ALMA maps originating at larger heights, it is valuable to  \hea{point out that there are no $T_\mathrm{b}$ peaks at locations with absolute LOS magnetic field strength larger than $\sim250$\,G, which suggests} that the presence of magnetic fields impacts the occurrence of the analysed dynamic small-scale events.

\subsection{Statistical study of \hea{detected} events}
\label{sec:Statistical study of events}

\begin{figure*}[tp]
\includegraphics[width=\textwidth]{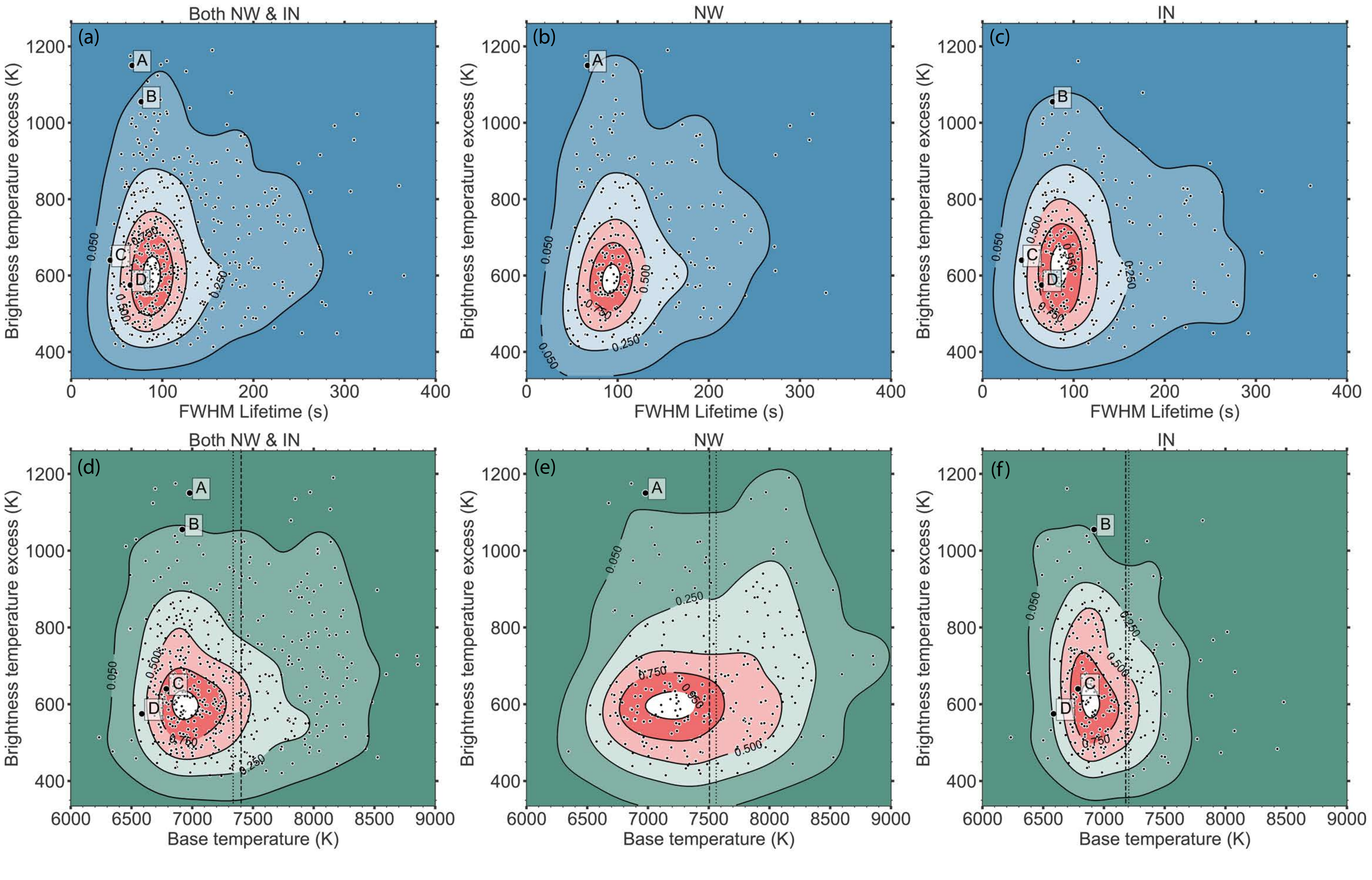}\caption{Statistical study of all the events with comparison between regions defined by the network internetwork mask. The first column (panels a $\&$ d) shows all the events, the second column (panels b $\&$ e) shows the events within the network mask and the third column (panels c $\&$ f) shows the events within the internetwork mask. For each event, the first row show the brightness temperature excess plotted against the lifetime and the second row show the brightness temperature excess plotted against the base brightness temperature. The distributions are represented as density maps and the contour lines marks the levels of $0.05$, $0.25$, $0.5$, $0.75$ and $0.95$. The dotted and dashed lines in panels d,e and f shows the median and mean value respectively of the brightness temperature of all pixels withing $15"$ during the entire dataset. The selected events A-D (Sect.\ref{sec:event_detailed_study}) are marked in all panels for reference.}
\label{fig:peak_lifetime_mag}
\end{figure*}

\hea{For each of the $552$ events (described in Sect. \ref{sec:results:peak_clustering}) the maximum $T_\mathrm{b}$ excess, lifetime and base brightness temperature (cf. Fig.\,\ref{fig:shockillustration}) are given in Fig.~\ref{fig:peak_lifetime_mag}.}
As mentioned in Sect.~\ref{sec:temp_var_obs_data}, the $T_\mathrm{b}$ values in the outskirts of the FOV are less certain 
\hea{and events detected there are therefore not considered further for the detailed analysis of event properties.}
In the panels of the first column, all the events are plotted while the events are split up according to the NW/IN mask in the second (NW) and third (IN) column. In all panels, the selected examples A -- D are marked, which are described in detail in Sect.~\ref{sec:event_detailed_study}. 
In the first row, the lifetime of the events is plotted against their maximum brightness temperature excess ($\Delta T_\text{b}$). 
While the distribution in both lifetime and $\Delta T_\text{b}$ appears to be continuous for all the events (Fig.~\ref{fig:peak_lifetime_mag}a), there are a couple trends worth noticing:

$(i)$~The density distribution of the scatter plot indicates the most frequent events have a typical brightness temperature excess of $\Delta T_\text{b} \sim 600$\,K  and a lifetime of $\sim90$\,s, with values between $\sim450$ -- $750$\,K and $\sim 55$ -- $125$\,s for $50~\%$ of the distribution.

\begin{figure}[tp]
\includegraphics[width=\columnwidth]{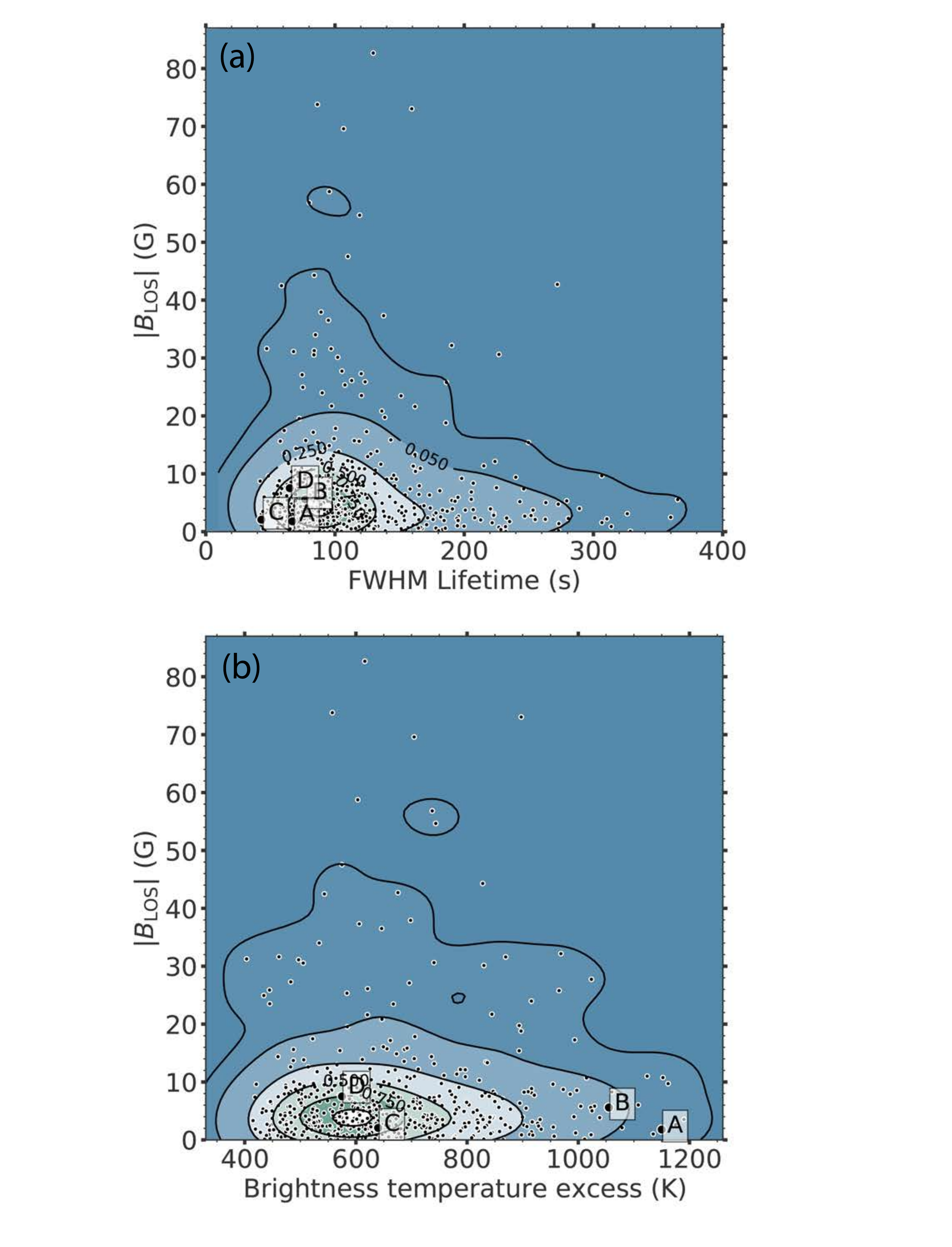}
\caption{Statistical study of all the events within $15\arcsec$ radius from the centre FOV. \textbf{a)} The absolute LOS magnetic field strength plotted against the lifetime for each event. \textbf{b)} The absolute LOS magnetic field strenght plotted against the $T_\mathrm{b}$ excess. The scatter plots are represented as coloured density maps with the contour lines marking the levels of $0.01$, $0.05$, $0.25$, $0.5$, $0.75$ and $0.95$ of the total distribution. The selected events A-D (Sect.\ref{sec:event_detailed_study}) are marked in all panels for reference.}
\label{fig:peak_mag_FWHMlifetime}
\end{figure}

$(ii)$~There are no events with a lifetime below $43$ s that also have a brightness temperature excess above $400$\,K. 
\swedt{This finding is discussed in connection with the effects of limited angular resolution in Sect.\,\ref{sec:disc-3D-simulations}.}
The events that fall within the network and the internetwork areas of the mask shown in Fig.~\ref{fig:peak_lifetime_mag}b and Fig.~\ref{fig:peak_lifetime_mag}c, respectively. The distributions of lifetime and brightness temperature excess for the network and the internetwork areas \hea{are} \swedt{qualitatively} \hea{similar} \swedt{but differ in a few aspects.} 
\hea{Most of the relatively few events with larger $T_\mathrm{b}$ excess (above $\sim1050$\,K} \swedt{are detected} \hea{in the NW regions. However, this does not imply that stronger magnetic field gives rise to larger $T_\mathrm{b}$ excess, as we will see in Fig.\,\ref{fig:peak_mag_FWHMlifetime}. Similarly, the events with low $T_\mathrm{b}$ excess and long lifetime (up to $\sim300$\,s) 
}\swedt{tend to} \hea{lie within the IN regions
}\swedt{but this affects only a small number of events}.

The base temperature (cf. Fig.\,\ref{fig:shockillustration}) could potentially affect the $T_\mathrm{b}$ excess amplitude of an event.
The brightness temperature excess is plotted against the base brightness temperature for all events in Fig.~\ref{fig:peak_lifetime_mag}d. There is a continuous distribution of events \hea{with} base brightness temperatures between $\sim 6400$ -- $8300$\,K for all brightness temperature excess values $\Delta T_\text{b}$ between $\sim400$ -- $1000$\,K. The density plot in Fig.~\ref{fig:peak_lifetime_mag}d indicates that most events exhibit a base brightness temperature of around $6600$ -- $7500$\,K. 
The mean and median brightness temperatures of all pixels within the same area \hea{as the considered events} ($15\arcsec$ radius \hea{and NW/IN respectively}) for the entire time of the dataset are marked as references in Figs.~\ref{fig:peak_lifetime_mag}d -- f. 
\swedt{The mean and median temperatures of the whole area are} $<T_\text{b}>_{x,y,t} = 7402$\,K \hea{and $7336$\,K,} \swedt{respectively,  which corresponds to} the upper segment of the base temperatures shown for the events.
This is naturally a cause of the base temperatures being defined at local minima \hea{of the $T_\mathrm{b}$ evolution}.
The separation of the events in NW and IN areas shows that there are more events with higher base temperature in the NW areas than in the IN areas (Figs.~\ref{fig:peak_lifetime_mag}e -- f). In accordance to higher (average) temperatures in the NW regions, the majority of events in the NW areas exhibit base temperatures of up to $\sim8100$\,K whereas there only few events in the IN areas with such high values. The lower limit is however almost the same for the two masked areas, \hea{which comes naturally as the NW mask contain very quiet patches as well at some locations and times}. 

For the NW regions alone, the mean and median values are $7561$\,K and $7506$\,K, and for IN alone the values are smaller, $7204$\,K and $7179$\,K, respectively. The mean of the brightness temperature is slightly higher than the median, most pronounced in the NW areas but also evident in the IN areas, which suggests that there are a few but very bright features. \hea{Even though there is a significant shift towards larger base brightness} \swedt{temperature} \hea{
for the NW regions, the distribution of magnitude of the $T_\mathrm{b}$ excess is similar and there is only} a mean brightness temperature difference of about $360$\,K between NW and IN areas. These results imply that the \hea{physical conditions} behind the occurrence of \hea{the registered} dynamical $T_\mathrm{b}$ excess events are not very dependent on base temperatures in the analysed quiet Sun region.

As expected from the studies of other chromospheric diagnostics \citep[see, e.g.,][and references therein]{
2009SSRv..144..275D,2009A&A...497..273L}, the (base) brightness temperature in ALMA Band~3 correlates with the (time-averaged) magnetic field strength in the photosphere \citep[see][for the analysed data set]{2020A&A...635A..71W}. 
\hea{In Fig.~\ref{fig:peak_mag_FWHMlifetime}, \hea{for each of the $552$ events,} the relations between the magnitude of the maximum brightness temperature excess $\Delta T_\text{b}$, the lifetime and the SDO/HMI \hea{LOS} magnetic field strength are shown. The absolute LOS magnetic field strength is given at the spatial location and the time of the maximum excess $T_\mathrm{b}$ peak for each event.}
The absolute magnetic field strength at the locations of the detected events ranges \hea{up to to $\sim80$\,G}.
The \hea{majority} of the events occur at locations with very low magnetic field strength below $10$\,G and as many as $95$\,\% of the events are found at locations with an absolute magnitude of $\leq 20$\,G.
There are some events with higher absolute magnetic field strength which are primarily occurring in the network areas. However, even in the network areas, the locations of a majority of the events show a low magnetic field strength similar to the events in the internetwork areas. This is because of small scale \hea{structure of} the magnetic field that \hea{the time-averaged NW/IN} mask does not account for. The network regions therefore still contain many areas with low average magnetic field strength, as can be seen in Fig.\,\ref{fig:mag_cont_mask}a, \hea{where there are many white patches also in the NW area.}
The events associated with larger absolute \hea{LOS} magnetic field strength have \hea{shorter lifetimes, centered around $100$\,s, similar to the majority of events.} \hea{The events with longer lifetimes tend to} \swedt{occur at location with} \hea{a weak magnetic field.}
\hea{A similar trend, but not equally clear can be seen in the brightness temperature excess with a larger spread of values for the events with larger magnetic field strength, similar to the distribution of the events with weak magnetic fields.}
These results could suggest that a strong \hea{LOS} magnetic field disturbs the generation of \swedt{events}  \hea{with long lifetime} or higher amplitude ($\Delta T_\text{b}$).
\hea{Even though the statistical sample is rather large with $552$ events, the number of events with larger magnetic field strength should be larger to draw conclusions about this.}

\begin{figure}[tp!]
\includegraphics[width=\columnwidth]{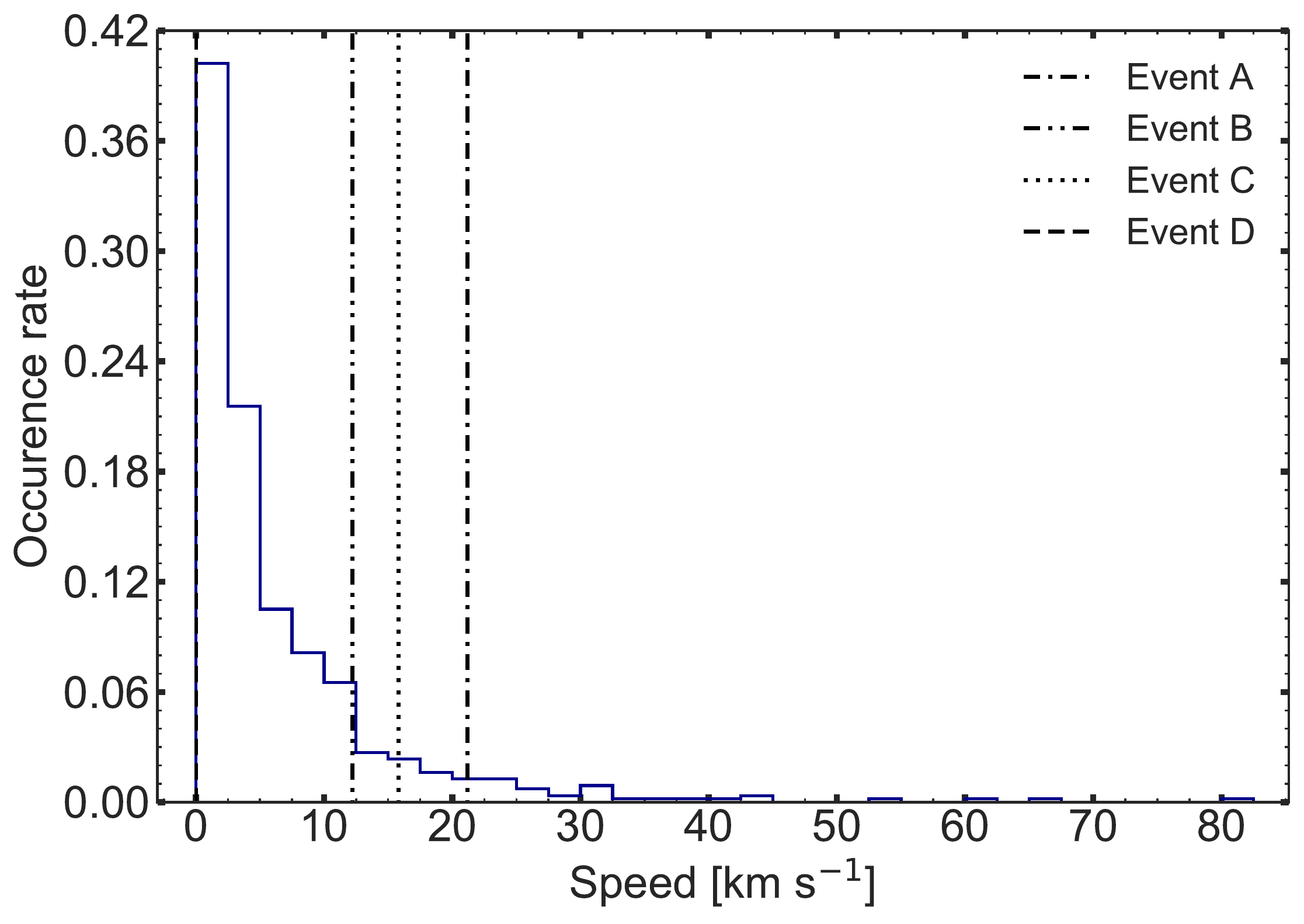}\caption{Distribution of the speed of apparent motion of the 552 events. The speeds of event A -- D (Sect.\ref{sec:event_detailed_study}) are marked for reference.}
\label{fig:apparent_motion}
\end{figure}

\paragraph{Apparent motion.}
In addition, the apparent motion is estimated for each event. This is done by looking at the change of position of the brightest point within the spatial extension of the event between the points in time at FWHM of the peak with largest $T_\mathrm{b}$.
The individual velocity components in x- and y-axis are found to be normally distributed around zero, which insinuates that there is no preferred direction of apparent motion of the events. The probability density function of the total absolute apparent velocity is given in Fig.\,\ref{fig:apparent_motion}. Most of the events, approximately $80$\,\%, show low speeds below $10$\,$\text{km\,s}\,^{-1}$ and the rest show speeds between $10$ -- $30$\,$\text{km\,s}\,^{-1}$,
except for a handful of events with indications of high speeds of several tens km\,s\,$^{-1}$.

\hea{\paragraph{Repetitiveness of events.} Some of the events occur on similar locations but with a delay in time, which can be inferred from comparing and tracing individual events between the panels in Fig.\,\ref{fig:events_coordinates}. An estimation of the repetitiveness of events is made by taking the spatial and temporal locations of the points with largest magnitude of $T_\mathrm{b}$ excess in each event and calculate how many of these are within a certain spatial distance from each other. Approximately $22\%$ of the events occur within a spatial distance of $1\arcsec$ and within the same time scan of $\sim10$\,min (Sect.\,\ref{sec:methods_obs_data}) as another event. For these events occurring at close-by locations,}
\swedt{the delay between the occurrence of the brightness temperature peaks of two events is typically} \hea{$\Delta t_\text{peaks}=200$\,s} \swedt{although a continuous distribution of delays of} \hea{up to approximately $\Delta t_\text{peaks}=320$\,s} and in a few cases even up to $\Delta t_\text{peaks}=430$\,s is found. It should be noted that the statistical significance of the derived delays is limited by the duration of the observational time scans and the spatial threshold (here set to $1\arcsec$) in relation to the angular resolution.
\hea{The complexity of} the 3D structure of the solar chromosphere on small spatial scales in combination with the limited angular resolution 
\swedt{can significantly affect the observable signatures of dynamic events at mm wavelengths (see the discussion in Sects.~\ref{sec:disc-3D-simulations}-\ref{sec:disc-spatial resolution}). Consequently, propagating shock waves that are } \hea{excited at photospheric heights can look very different even if they would be excited at the same location.  
Indications of the repetitiveness of events can be seen in the time evolution graphs of the selected examples in Figs.\,\ref{fig:custom2} -- \ref{fig:custom_s2} and Figs.\,\ref{fig:custom1} -- \ref{fig:custom_sm}.}

\subsection{In-depth study of selected events in the observational data} 
\label{sec:event_detailed_study}
\begin{figure*}[tp!]
\includegraphics[width=\textwidth]{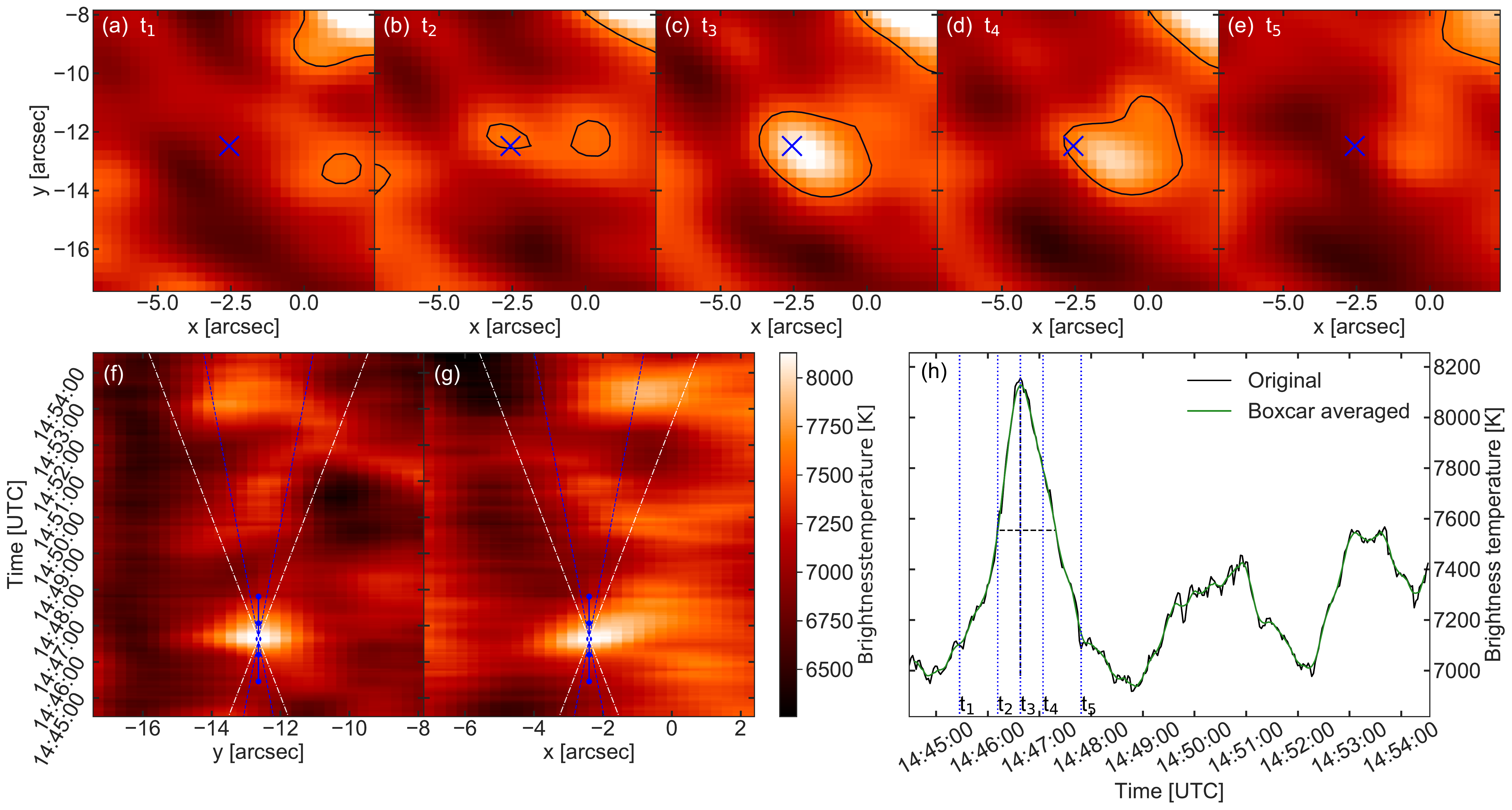}
\caption{Detailed study of event A.  
\textbf{a-e)} The top panels show a close up of the surroundings at different time steps, $t_1$ to $t_5$ from left to right, through the shock wave event. The time of the $T_\mathrm{b}$ peak is marked by $t_3=14$:$46$:$38$. $t_1$ and $t_2$ mark $70$\,s and $26$\,s prior the peak and $t_4$ and $t_5$ mark $26$\,s and $70$\,s after the peak. The center location is marked by a blue cross. The contour lines marks the half maximum of the maximum $\Delta T_\text{b}$ peak. 
\textbf{f-g)} Space-time diagrams for a vertical and horizontal slit across the FOV at the center location, which is marked with blue dots for the time steps $t_1$ to $t_5$. Velocity slopes for $10$ and $20$\,km\,s\,$^{-1}$ are indicated by blue dotted and white dashed lines, respectively. The color code is the same in all panels. 
\textbf{h)} The time evolution of the brightness temperature of the center location, where the time steps $t_1$ to $t_5$ are indicated by blue dotted vertical lines. Both the original data (black) and the averaged data (green) are shown. The horizontal and vertical black dashed lines mark the event lifetime and brightness temperature excess.}
\label{fig:custom2}
\end{figure*}

\begin{figure*}[tp!]
\includegraphics[width=\textwidth]{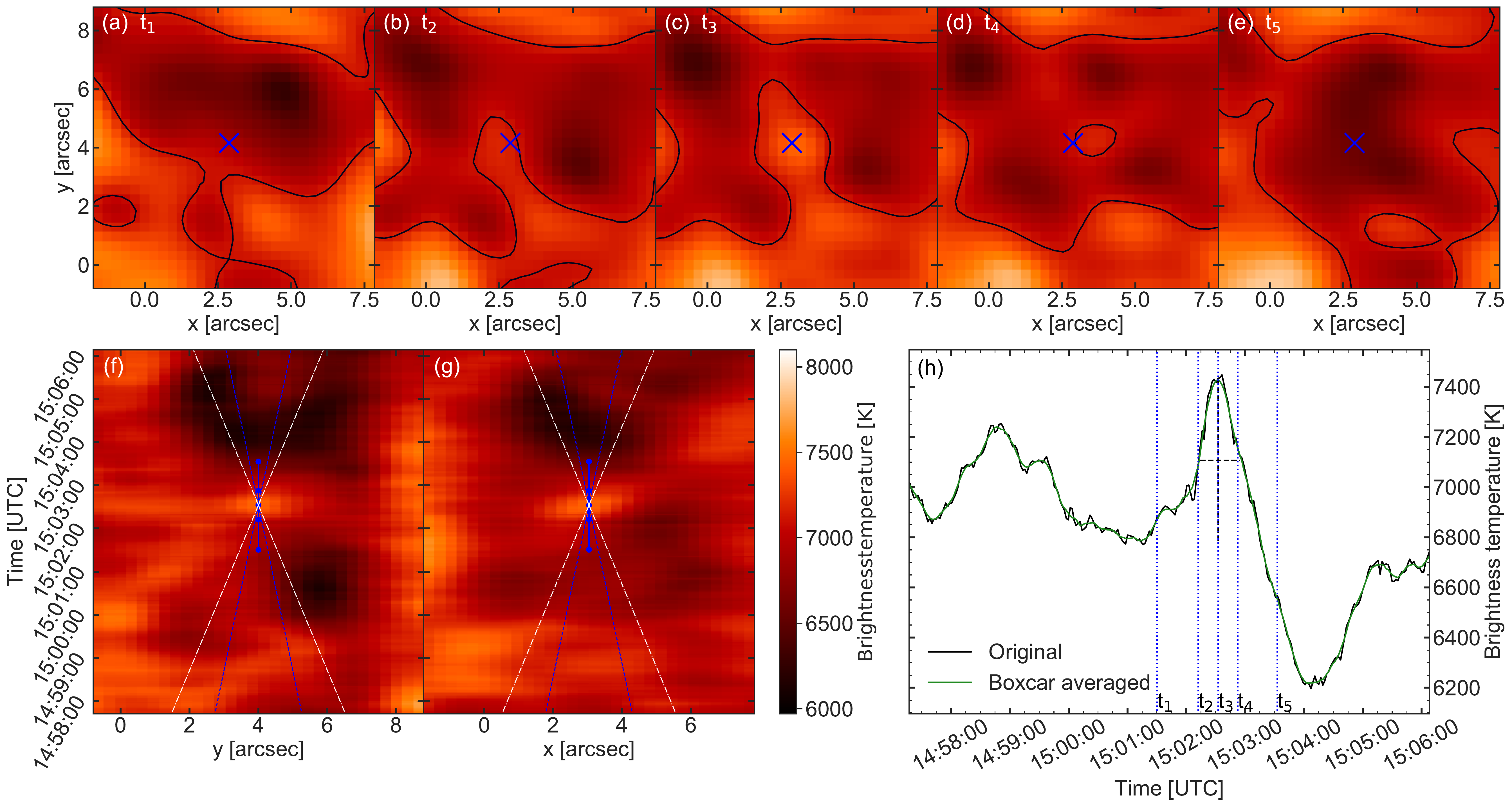}
\caption{Detailed study of event C. Description as for event A in Fig.\,\ref{fig:custom2}. The time of the $T_\mathrm{b}$ peak is marked by $t_3=15$:$02$:$32$. $t_1, t_2, t_4$ and $t_5$ mark $-60,-20,20$ and $60$\,s from the peak, respectively.}
\label{fig:custom_s2}
\end{figure*}

\hea{A couple examples of dynamic events, representative for different segments of the parameter space 
(see Figs.\,\ref{fig:peak_lifetime_mag} -- \ref{fig:peak_mag_FWHMlifetime}) are illustrated in Figs.~\ref{fig:custom2} -- \ref{fig:custom_s2}. }
\swedt{Event} (A) exhibits more pronounced brightness temperature amplitude and \hea{the other (C) with a weaker $T_\mathrm{b}$ excess}, representative of the most frequently detected events (as seen in Fig.~\ref{fig:peak_lifetime_mag}a). 
\hea{An additional two examples (events B and D), showing similar signatures yet with different details, are presented in the Appendix\,\ref{sec:appendix:events}.} \hea{All the four} events are marked in Figs.\,\ref{fig:events_coordinates} -- \ref{fig:peak_mag_FWHMlifetime} for reference.

Event~A, which is located at $(x,y)=(-2.4\arcsec,-12.6\arcsec)$, is shown in Fig.\,\ref{fig:custom2}. The lower right panel (h) shows the time evolution of the brightness temperature at the fixed location with a rise from $T_\text{b}=6980$\,K to $8130$\,K in about $53$\,s. After reaching the peak, the brightness temperature decreases 
again to $6940$\,K in the course of $67$\,s, to only slightly below the initial $T_\mathrm{b}$ value. 
Event~A has thus temperature excess of $\Delta T_\text{b} = 1150$\,K and a lifetime of $67$\,s. The area connected to the event, e.g. that shows a brightness temperature of at least half maximum of the $\Delta T_\text{b}$ peak, has a round shape with a maximum diameter of roughly $4$ arcsec. The upper row with five panels shows a time series of close-ups of the surroundings of the event. The event develops rapidly at one location in the FOV and expands in all directions and then decrease in brightness whilst remaining fairly (horizontally) stationary apart from a rightward motion component. 
Even though the \hea{limited angular} resolution makes it challenging, the apparent movement through time of the brightest point, of approximately  $22$\,km\,s$^{-1}$, can be seen in the space-time diagrams in the lower left panels (Fig.\,\ref{fig:custom2}f-g). As indicated in Fig.\,\ref{fig:mag_cont_mask}, event A is located in the network region, but still shows a weak absolute magnetic field strength of only $\sim 18$\,G. \hea{In the $T_\mathrm{b}$ time evolution (Fig.\,\ref{fig:custom2}h), two weaker succeeding peaks can be seen approximately $4.5$\,min and $7$\,min after the strong peak of event~A. Even though these specific peaks might not fulfill the $\Delta T_\text{b}>400$\,K  requirement used in the selection of the events (Sect.\,\ref{sec:temp_var_obs_data}), the time delay is in line with the numbers presented in Sect.\,\ref{sec:Statistical study of events} for the repetitiveness of events.}  \hea{Another strong event (B), showing many similarities to (A), is for reference illustrated in Fig.\,\ref{fig:custom1}.}

Event~C (Fig.\,\ref{fig:custom_s2}) occurs at the location $(x,y)=(4.0\arcsec,3.0\arcsec)$ close to event B, but more than $6$ minutes later in time. The brightness temperature increases from $6790$\,K to $7430$\,K, resulting in $\Delta T_\text{b} = 640$\,K and a lifetime of $43$\,s. 
After reaching the $T_\mathrm{b}$ peak, there is a significant drop to about $6200$\,K. As can be seen in the space-time diagrams in the lower left panels, the temperature in the area drops to about $6000$\,K. We conclude that this more pronounced contrast is likely caused by the formation of a slightly larger than average post-shock region. In ALMA data presented here, such regions are typically seen as round shapes with diameters on the order of up to a few arcsec. Most of these regions, however, have typical diameters of $1-2\arcsec$, which is at or below the resolution limit in the ALMA Band~3 data used here. 
The post-shock region occurring in connection to event~C is slightly larger and therefore better resolved, thus revealing a larger $T_\mathrm{b}$ contrast. 
See \hea{Sects.\,\ref{sec:disc-3D-simulations} -- \ref{sec:disc-spatial resolution} \citep[and in addition, e.g.,][]{2007A&A...471..977W, 2015A&A...575A..15L}} for a discussion on the impact of low spatial resolution on observations of propagating shock wave patterns. 
The bright feature appears stable in the horizontal plane with just a subtle movement towards higher values on the x-axis (Fig.\,\ref{fig:custom_s2}. The speed of the apparent motion is $\sim 15$\,km\,s$^{-1}$ at FWHM. \hea{An event (D), similar to event~C, but supposedly even more on the limit of being resolved is presented in Fig.\,\ref{fig:custom_sm}.}

As we will discuss in Sect.~\ref{sec:discuss_resolution}, the angular resolution achieved for the analysed ALMA Band~3 observations is at the limit for resolving the chromospheric shock-wave-induced pattern. 
In areas with short distances between neighbouring shock waves, most of the resulting mesh-like pattern remains unresolved. There, the shock peaks do not stand out from the surroundings, but instead the observed  dynamical structure is dominated by the darker (post-shock) regions. \hea{The latter example, event~C, indeed shows less clearly isolated features (alike for event~D in Fig.\,\ref{fig:custom_sm})}. 
\hea{This could contribute to apparent lower values of the most represented $\Delta T_\text{b}$ amplitudes than what would be seen with higher angular resolution.}
%



\section{Discussion}
\label{sec:disc}

\subsection{\hea{1D numerical simulations of propagating shock waves}}\label{sec:disc-simulations}

In order to check if propagating shock waves could be producing the observed brightness temperature events, we study numerical simulations and calculate the radiative transfer in mm-wavelengths.
\hea{The hydrodynamic code RADYN \citep{1992ApJ...397L..59C, 1994chdy.conf...47C, 1997ApJ...481..500C, 2002ApJ...572..626C} produces one-dimensional dynamical model atmospheres where a moving piston in the lower boundary excites waves. RADYN solves the equations of conservation of energy, momentum and mass, the radiative transfer equations in non-local thermodynamic equilibrium and the rate equations of atomic level populations.}
\hea{\cite{2004A&A...419..747L, 2006A&A...456..713L} present  $T_\mathrm{b}$ signatures at  mm wavelengths resulting from   RADYN runs for which  the waves are excited by the driving piston} \swedt{with velocities set to match photospheric Doppler velocities obtained from observations in the \ion{Fe}{I}} \hea{line at $\lambda=396.68$\,nm   \citep{2002ApJ...572..626C}. 
In the time series with length of $\sim1$\,h, waves with varying strength and speed are excited, allowing some of them to interact. The resulting shock wave signatures at} \swedt{a wavelength of $3.4$\,mm wavelength (see Fig.~1c in \citeauthor{2006A&A...456..713L}~\citeyear{2006A&A...456..713L}), which is close to the wavelengths currently offered for solar observing with ALMA in Band~3,}  \hea{show $\Delta T_\text{b}$ amplitudes between a few hundred K to $\sim4000$\,K and lifetimes in the span between approximately $15-50$\,s. Some events are double-peaked, showing a long lifetime, which seems to be the result of interacting waves.} Wave interference can thus significantly contribute to extending the apparent lifetime of an event.
\hea{To be unbiased of specifics of the input observational Doppler velocities, we perform a parameter exploration with series of monochromatic waves covering large parameter ranges. This allows to see 
under which circumstances shock waves
could give rise to T$_\text{b}$ signatures that are in line with observed signatures described in Sect.\,\ref{sec:Statistical study of events}.}
\hea{The RADYN code is used and the initial conditions are set by a VAL model atmosphere \citep{1981ApJS...45..635V} with the addition of a transition region and corona \citep{2002ApJ...572..626C}.}
The lower boundary is set to $100$\,km below where the optical depth is unity at a wavelength of $500$ nm. The upper boundary is at a height of $10^4$\,km, where the temperature is set to $10^6$\,K. To resolve small scale variations arising from the propagation of shock waves, an adaptive mesh is used \citep{1987JCoPh..69..175D}. The duration of the simulations is $1500$ seconds and thus long enough for the relaxation phase of the atmospheres to pass and for several shock waves to propagate through. The cadence of the output is $1$\,s. \hea{The piston at the lower boundary excites sinusoidal acoustic waves where the initial amplitude and the periodicity are varied independently.} 

The amplitude is varied between $0.001$ and $0.15$ of the local sound speed, $v_0$, at the lower boundary and the periodicity between a few seconds up to several minutes. 
Distinct shock waves are excited with an initial amplitude between $0.005 v_0 - 0.05 v_0$ and wave periodicity between $P \approx 90 - 210$\,s. \hea{The waves steepens into shock waves around $1$\,Mm and} propagate through the chromosphere with a speed of approximately $9.4$\,km\,s$^{-1}$. The radiative transfer through the dynamical atmosphere models is calculated using the RH 1.5D code \citep{2001ApJ...557..389U, 2015A&A...574A...3P}. \hea{The output continuum intensities ($I_\nu$) are converted to brightness temperatures through relation\,(\ref{eq:tb}).}

\begin{figure}[tbh]
\includegraphics[width=\columnwidth]{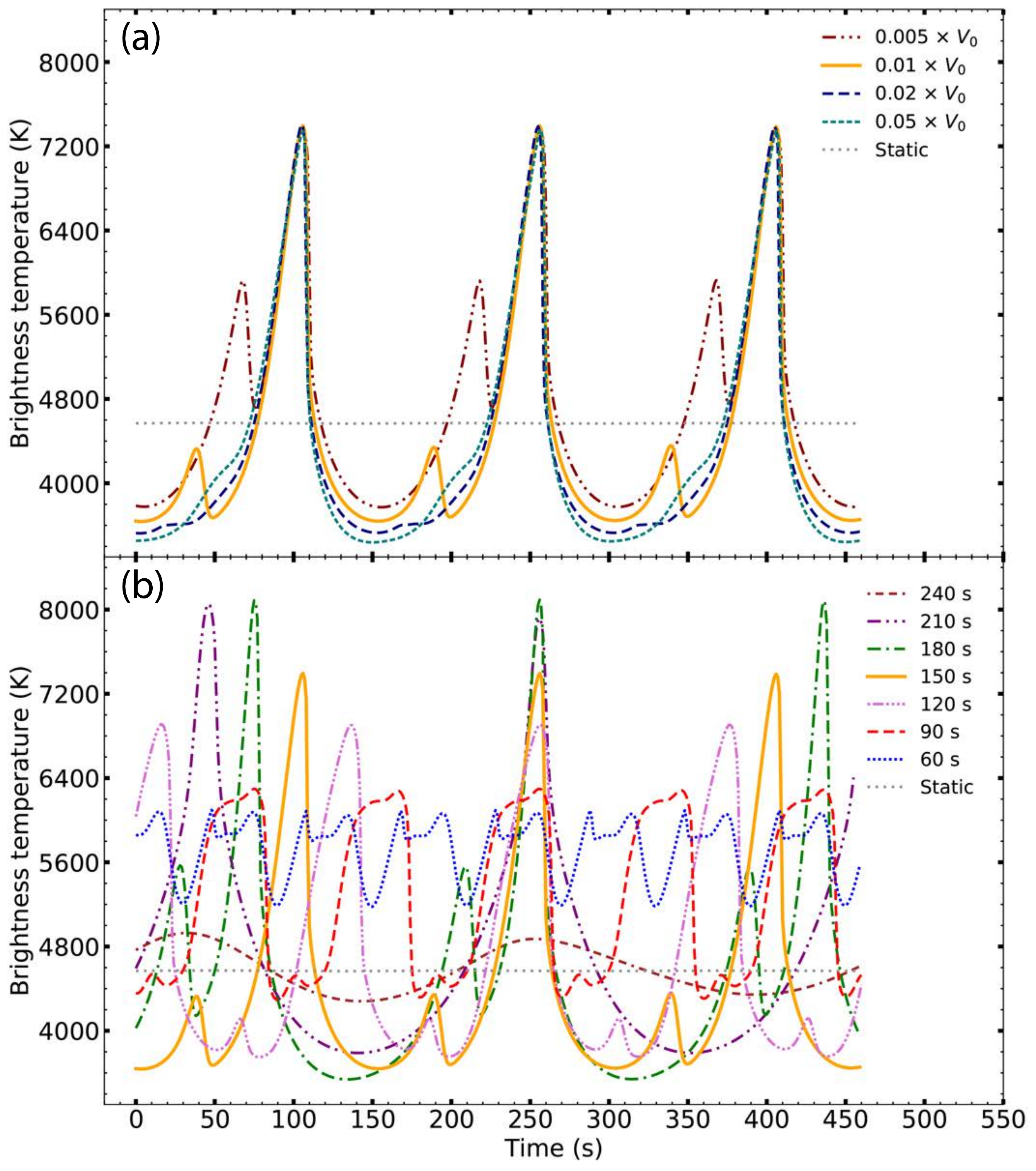}
\caption{Evolution of brightness temperature with propagating shock waves. \textbf{a)} Variation of the initial wave amplitude $A_i$ between $0.5 \%$ and $5 \%$ times the local sound speed at the lower boundary ($v_0$). \hea{The periodicity is kept at $150$\,s.} \textbf{b)} Variation of the wave periodicity between $60$ -- $240$\,s, where $A_i$ is kept at $0.01$\,$v_0$. The orange graph in panel a and b thus represents the same run. In both panels, the case of a static atmosphere without any disturbances is plotted as reference with a grey dotted line. The runs are shifted in time to align the time of their peaks.} 
\label{fig:temp_peaks}
\end{figure}

For each time step, the mean value of the brightness temperatures of all four sub-bands of ALMA band~3, see Table \ref{tab:frequencies}, is calculated to match the spectral setup of the full band integrations in the ALMA observations, \hea{which makes a one-to-one comparison possible.} The temporal evolution of the simulated Band~3 brightness temperature is shown for the different simulation runs in Fig.~\ref{fig:temp_peaks}.
\hea{Each shock wave signature consists of a rapid increase in brightness temperature, reaching a peak and followed by a post-shock drop in brightness temperature. 
The case of a static unperturbed atmosphere is given as a reference.}

Worth noting is that, when going towards lower initial wave amplitudes, a preceding temperature peak is growing stronger and partly merges into the main peak. Presence of a preceding peak is most prominent in the case of $A_i=0.005 v_0$.
where the minimum brightness temperature is significantly enhanced in connection to the main peak for more than $60$\,s, as well as throughout the entire cycle. 
\hea{Such double-peaked temporal profiles are also seen in the series with different wave modes \citep{2004A&A...419..747L, 2006A&A...456..713L}, as a result of interference between waves. We also note that some of the observational examples, specifically event~C (Fig\,\ref{fig:custom_s2}), exhibits a very subtle increase of $T_\mathrm{b}$ preceding the main peak, which might be consistent with the small preceding secondary peaks seen in the simulations for certain parameters.}

\begin{figure}[tbh]
\includegraphics[width=\columnwidth]{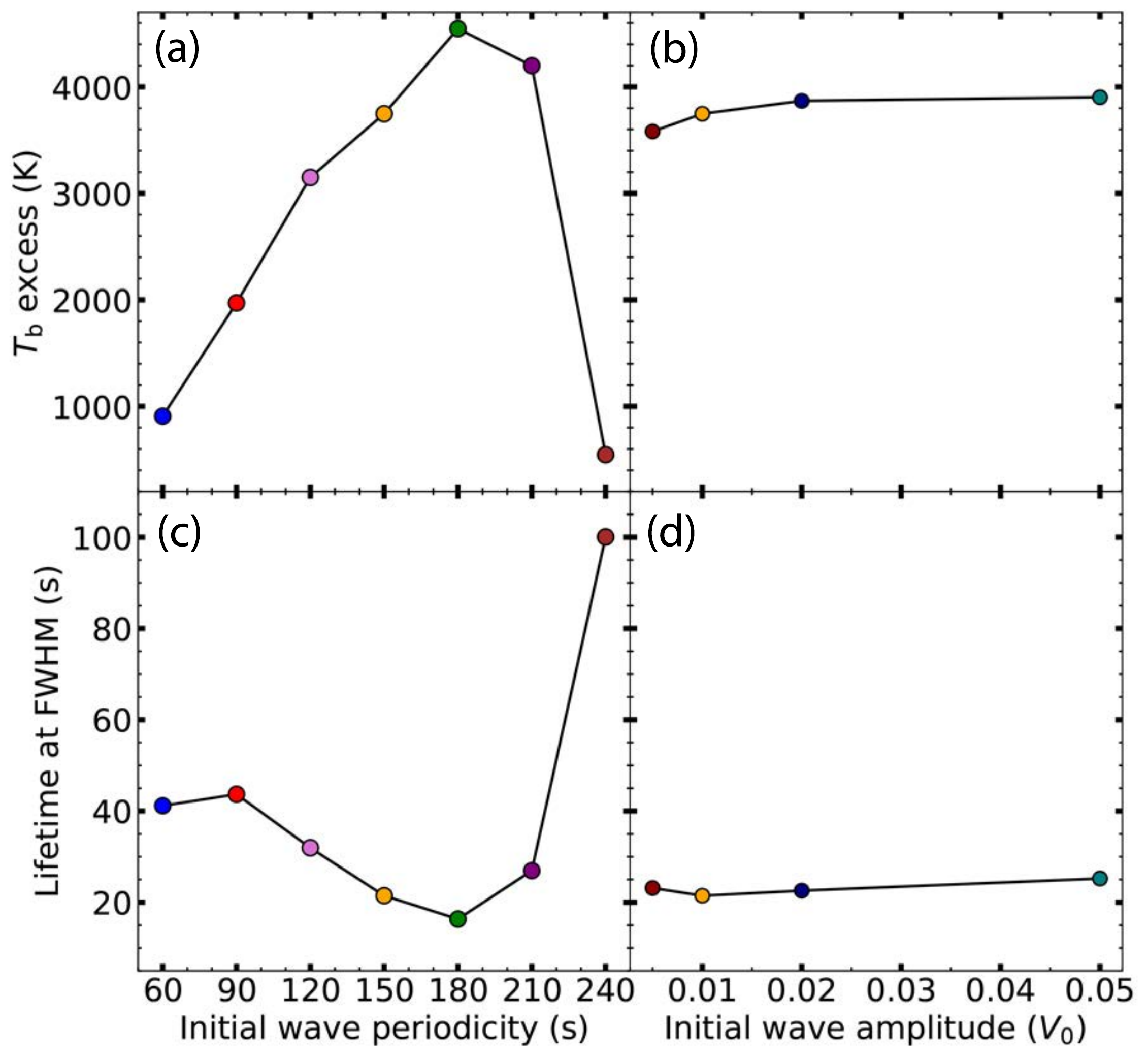}
\caption{Change of temperature and lifetime of shock wave signature with varying wave parameters. \textbf{a)}~Change of temperature as function of initial wave frequency. \textbf{b)}~Change of temperature as function of initial wave amplitude. \textbf{c)}~The lifetime at FWHM of the brightness temperature excess peak as function of initial wave frequency. \textbf{d)}~The $T_\mathrm{b}$ peak lifetime as function of initial wave amplitude. The color-scheme is the same as in Fig.\,\ref{fig:temp_peaks}.}
\label{fig:wave_parameter_exploration}
\end{figure}

Figure~\ref{fig:wave_parameter_exploration} shows the brightness temperature excess and the lifetimes of the shock wave signatures.
\hea{The parameter exploration reveals that, the} variations of the initial amplitude only have a small impact on both the brightness temperature excess \hea{$\Delta T_\text{b}$}, which is found to be on the order $\sim3500$ -- $3900$\,K and on the lifetimes, with values between $21$ -- $25$\,s.
\hea{In contrast, the wave periodicity has a large impact on the brightness temperature excess and signature lifetime (Fig.~\ref{fig:wave_parameter_exploration}, panel a and c). In the case of long periodicity of $P=240$\,s, the $T_b$ signature shows a low brightness temperature excess of only about $\sim 550$\,K with a corresponding long lifetime of $\sim100$\,s. This agrees with the results of \cite{2006A&A...456..713L}, where they show that at wavelengths of ALMA band 3, the intensity in the power spectrum at $240$\,s ($4.2$\,mHz) is very low whilst the peak is found at a period of $175$\,s ($5.75$\,mHz), which agrees well with Fig.\ref{fig:wave_parameter_exploration}a. A short periodicity of $P=60$\,s, gives rise to $\Delta T_\text{b} = 900$\,K and a lifetime of $40$\,s. However the bulk shift towards higher temperatures (Fig.\ref{fig:temp_peaks}b), indicates that the atmosphere is not allowed to relax between the passages of waves.}
\hea{For the cases between $P=90$\,s to $210$\,s, with clear $T_\mathrm{b}$ shock wave signatures, values of the brightness temperature excess between $\sim2000$\,K to $4450$\,K are seen, together with lifetimes between $16$\,s -- $45$\,s. This is consistent with the results from the series of waves derived from photospheric Doppler velocity measurements \citep{2006A&A...456..713L}.}

\hea{In conclusion, shock waves with reasonable wave parameters, which are in agreement with observational constrained values, give rise to signatures with brightness temperature excess $\Delta T_\text{b}$ from a few hundred K up to order of $4500$\,K  (at wavelengths of ALMA band\,3). $\Delta T_\text{b}$ of $4500$\,K are about $3$\,--\,$4$ times larger than what is seen in the observational ALMA data. In the following section we explore a possible explanation for this.}

\subsection{\hea{Shock wave propagation in a 3D environment}}\label{sec:disc-3D-simulations}

The one-dimensional simulations are only an idealized first approximation, whereas for shock wave signatures in the observational data the following two factors have to be considered: 
$(i)$~Many of the shock wave signatures are not fully resolved and there is spatial smearing with the surroundings that comes from sampling with the relatively large clean beam. 
\cite{2006A&A...459L...9W} demonstrated that the dynamical time scale of the chromospheric small-scale internetwork structure does indeed increase when lowering the spatial resolution. 
$(ii)$~The shock waves propagate through an inhomogeneous three-dimensional medium that is shaped by previous waves and magnetic fields. The shock waves can also interfere with each other and an fully isolated shock wave event as in the one-dimensional simulations might be difficult to observe. Both constructive as well as destructive interference could change the observed brightness temperature.

\hea{Propagation of shock waves in a more realistic 3D environment} 
\swedt{is demonstrated here means of numerical simulations with the Bifrost code 
\citep{2011A&A...531A.154G,2016A&A...585A...4C}. 
See \citet{2020arXiv200805324E} for a detailed analysis of an exemplary shock wave event. 
Here, we study the observational imprint of}  \hea{shock wave propagation in brightness temperature maps calculated with the Advanced Radiative Transfer (ART) code (de la Cruz Rodriguez~et~al., in prep.), at wavelengths corresponding to ALMA band 3 (cf. Table\,\ref{tab:frequencies}). The ART code assumes LTE but takes into acccount in detail relevant sources of continuum opacity.}

\begin{figure*}[tbh]
\includegraphics[width=\textwidth]{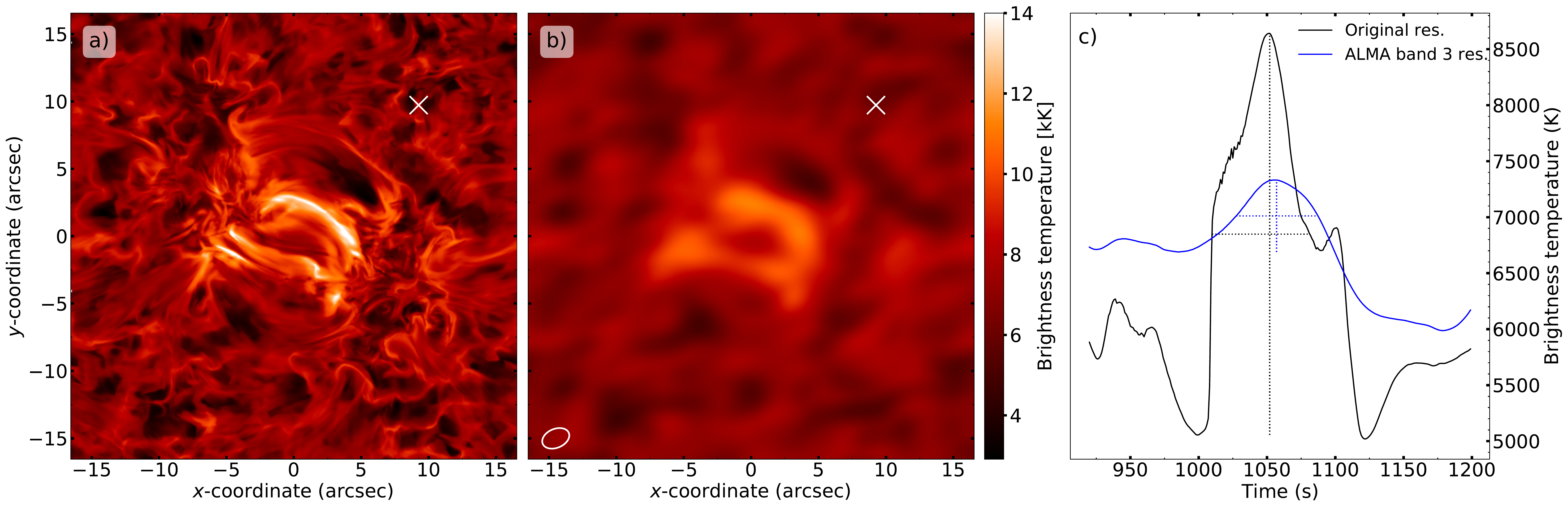}
\caption{\hea{Example of degradation with lower spatial resolution of shock wave $T_b$ signature from the 3D Bifrost numerical model.
\textbf{a)} Synthetic brightness temperature map with a cell size of $\sim0.066\arcsec$, averaged over the frequencies of ALMA band 3 at $t=1052$\,s. The white cross marks the location of the selected event, $(x,y)=(9.26,9.72)$\,Mm. 
\textbf{b)} The synthetic brightness temperature map degraded to the spatial resolution of the ALMA band 3 observations by convolution with the clean beam of observational data. The FWHM of the clean beam is marked by the white ellipse for reference.
\textbf{c)} Time evolution of the brightness temperature of the original and convolved maps at the selected location between $t=920$\,s and $1200$\,s. The vertical and horizontal dotted lines marks the $T_\mathrm{b}$ excess amplitude and lifetime, of the shock wave signature in respective resolution. }} 
\label{fig:bifrost_ex}
\end{figure*}

\hea{To estimate the magnitude of the degradation that comes with limited spatial resolution, the maps are degraded to the resolution of the ALMA band 3 observations by convolution with the clean beam (Sect.\,\ref{sec:methods_obs_data}). }
\swedt{An original $T_\mathrm{b}$ map for a selected time step is compared to the corresponding spatially degraded map in Fig.\,\ref{fig:bifrost_ex}a-b). 
As expected, the fine structure on spatial scales smaller than the resolution limit is lost in 
}\hea{
the spatially degraded map. 
The location of an example of a propagating shock wave is marked in respective mm-maps. The evolution of the gas temperature surrounding the shock front is for reference given in Fig.\,\ref{fig:appendix:bifrost}. The time evolution of the brightness temperature at that location is shown in Fig.\,\ref{fig:bifrost_ex}c for both the original map and the spatially degraded map.  The propagating shock wave gives rise to a signature with a magnitude $\Delta T_\text{b}=3600$\,K, in the map with original resolution. In the map with degraded spatial resolution, the shock wave shows a signature with a substantially reduced magnitude of  $\Delta T_\text{b}=640$\,K. The limited spatial resolution of the ALMA band\,3 observations thus results in an observable signature of this shock wave example, $\sim0.18$ times the original $\Delta T_\text{b}$ magnitude.
Applying this conversion factor to the temperature excess seen in the 1D simulations with values between $\sim2000$\,K -- $4450$\,K,} \swedt{would result in reduced values of} \hea{$\sim360$\,K -- $800$\,K. These values agrees very well with what is seen in the observational data (Fig.\,\ref{fig:peak_lifetime_mag})}
\swedt{and thus implies that the observed signatures could be caused by} 
\hea{propagating shock waves.}

\hea{This is also in line with what \cite{2020A&A...635A..71W} demonstrated using the same Bifrost model. They find that applying} the Band~3 synthetic beam reduces the brightness temperature standard deviation $T_\text{b}^\mathrm{rms}$ from $1794$\,K to $1254$\,K for network pixels and from $1304$\,K to $693$\,K for internetwork pixels, respectively, 
\hea{i.e., a factor of $0.70$ for NW and $0.53$ for IN.}
Applying these factors to the range of $900$\,K to $4600$\,K found with the 1D RADYN/RH simulations in Sect.~\ref{sec:disc-simulations}, reduces the expected range of brightness temperature amplitudes in shock waves to $630$\,K -- $3220$\,K for network pixels and to $480$\,K -- $2440$\,K for internetwork pixels.  
We note that interpolating the resolution-dependent $T_\text{b}^\mathrm{rms}$ values by \citet{2015A&A...575A..15L} to the resolution of the Band~3 data (here $1.4$\arcsec -- $2.1$\arcsec) also result in a reduction by a factor $\sim 0.7$ with respect to the original resolution.
\citet{2007A&A...471..977W} also investigated the impact of the spatial resolution on their simulated brightness temperature maps showing a small-scale shock wave pattern, although for a shorter wavelength of 1\,mm. 
Using their Eq.~(9) with a characteristic length scale of $D = 1000$\,km and the angular resolution of the 
Band~3 data used here ($2.1\arcsec$ -- $1.4\arcsec$) results in a reduction by a factor $0.22$ -- $0.36$. 
The expected range of shock wave brightness temperature amplitudes based on the RADYN/RH results presented here would then be reduced to a lower limit of $200$\,K -- $330$\,K and an upper limit of $1000$\,K -- $1700$\,K, which is much more in line with the range of the values found in the observations in Sect.~\ref{sec:Statistical study of events} (see Fig.~\ref{fig:peak_lifetime_mag}).

\hea{The lifetime of the shock wave example is in original resolution $70$\,s, and $61$\,s in the degraded resolution (Fig.\,\ref{fig:bifrost_ex}c). The cooler pre-shock period around $t=1000$\,s, is not apparent in the ALMA band 3 resolution which contributes to}
\swedt{a shorter apparent lifetime. This example illustrates that the}
\hea{lifetime of an event is heavily dependent on the surrounding dynamical structure and the extent (and orientation) of spatial smearing due to limited angular resolution.}
\swedt{The resulting longer lifetimes seen in the 3D model and the observations as compared to the shorter lifetimes in the 1D simulations (Figs.\,\ref{fig:temp_peaks}--\ref{fig:wave_parameter_exploration}), 
emphasises the short-comings of the 1D approach in this particular aspect and the need for a systematic study based on 3D simulations.}  

The 1D RADYN simulations predict lifetimes for shock wave signatures in the range from $15$\,s to $50$\,s (see Sect.~\ref{sec:disc-simulations}). The simulations also show that it is possible that hydrodynamical wave motions, although not steepening into shock waves, could show longer lifetimes on the order of $\sim100$\,s with a brightness temperature excess of at least $400$\,K  (Fig.\,\ref{fig:temp_peaks}). The simulations indicate short-lived signatures well under one minute for shock waves. However with a potential broadening of the temporal profiles, longer lifetimes are expected. How much this broadening could be\hea{,} is strictly dependent on the spatial smearing \hea{of the radiative maps}, i.e., the evolution of the surroundings of the signatures and the size of the clean beam. 

While it is evident that the observable signatures in Band~3 brightness temperatures are highly resolution-dependent, it is also clear that a detailed comparison of simulated and observed signatures and a possible identification of shock waves as their source require a more direct approach using detailed three-dimensional (3D) simulations (see Sect.~\ref{sec:discuss_1dvs3d}). 

\subsection{\hea{Detection of events with limited spatial resolution}}
\label{sec:disc-spatial resolution}

Even though many dynamical events with brightness temperature amplitudes of at least $400$\,K are detected, 
the distribution of the spatial sizes of the events (see Fig.\,\ref{fig:size_histogram}) is most likely cut-off at spatial scales corresponding to the size of the clean beam. It is plausible that the distribution can be extrapolated towards smaller sizes and that smaller events thus remain unresolved.

This assumption is supported by \hea{the study of the Bifrost simulation (Fig.\,\ref{fig:bifrost_ex}),} observations in other chromospheric diagnostics, e.g. in the \ion{Ca}{II}\,K line at a spatial resolution of 0.7\arcsec \citep{2006A&A...459L...9W}, and other numerical simulations \citep[e.g.,][]{2004A&A...414.1121W} that clearly show small-scale features beyond the resolution of the here analysed ALMA Band~3 data. 
\label{sec:discuss_resolution}
Similar studies as this work using measurements with higher \hea{spatial} resolution would directly give insights to this. 
\hea{A higher spatial resolution, with the same wavelengths, could be achieved by making measurements with a more extended antenna array configuration.}

As mentioned in Sect.~\ref{sec:introduction}, the typical spatial scale of the mesh-like pattern in the chromosphere originating from propagating shock waves is of the same order as the angular resolution of the ALMA data used here \citep{2004A&A...414.1121W,2006A&A...459L...9W,2007A&A...471..977W,2015A&A...575A..15L}.
Consequently, the limited spatial resolution of this dataset ($1.4$ -- $2.1$ arcsec) does not allow to fully resolve the (sub)structure of the detected events, and potentially failing to detect small and weak events. 
Using the above-mentioned clustering technique, the $552$ detected dynamical events in this dataset (within $15"$ radius) must therefore be considered to be a lower limit only. 
Rather, only the strongest events and events with strong contrast with respect to their surrounding are detected. 
The effect of low angular resolution on the visibility of the chromospheric small-scale pattern is demonstrated by \citet{2007A&A...471..977W},  \citet{2015A&A...575A..15L}, and \cite{2020A&A...635A..71W}. 

The spatial resolution of interferometric data is often improved by the \textit{Earth-rotation aperture synthesis} technique \citep{1975Sci...188.1071R}. Although this is a genius technique, it is inherently limited to a temporal span below that of the typical dynamical timescale of the target, or else information will instead be lost. In the case of the highly dynamical chromosphere, the dynamical timescale is on the order of seconds. To properly resolve dynamical events we therefore require a cadence of a couple seconds at the most.

\subsection{Dependence on magnetic fields}

The results of this work show that the dynamical brightness temperature \hea{signatures are}  not evenly distributed in the FOV 
\swedt{but that strong magnetic fields coincide with a lower occurrence of} \hea{$T_\text{b}$ signatures (see Fig. \ref{fig:mag_cont_mask}).}
\hea{Despite the difference in formation heights between the (chromospheric) $T_\text{b}$ signatures} in the ALMA band~3 data and the \hea{(photospheric) LOS} magnetic field strength given by the SDO/HMI magnetograms, the lack of $T_\text{b}$ signatures, specifically at the footpoints \swedt{of stronger  magnetic field structures is seen. 
It should be emphasised again that it is important to specifically refer to the magnetic field strength at the exact time and location of the event, whereas the  network/inter-network mask shown in Sect.\,\ref{sec:magnetic field} \citep[cf.][]{2020A&A...635A..71W} is only providing a first reference for the overall properties of the magnetic environment.} 
\swedt{In summary, the lowered occurrence of events at times and locations of high magnetic field strength is in line with the observed signatures likely being caused by 
propagating shock waves.}

\subsection{Identification of shock waves.}
\label{sect:disc:Identify shock waves}
Based on the analysis of the mm-wavelength observables from the 1D numerical models (Sect.\,\ref{sec:disc-simulations}), we adopt here the following criteria for shock wave signatures.

\begin{enumerate}[(i)]
    \item A brightness temperature excess, $\Delta T_\text{b}$, \hea{on} order of a few hundred to thousands Kelvin.
    \item Lifetime \hea{on} the order of tens of seconds.
    \item Small lateral motion with speeds of less than a few tens\,km\,s$^{-1}$. 
    \item Low magnetic field strength.
\end{enumerate}

Taking a closer look at the selected events A\,-\,D in Sect.\,\ref{sec:event_detailed_study} reveals that they fulfill all of the conditions in the list above. The events show significantly large brightness temperature excess ($575$ -- $1150$\,K) as well as satisfying lifetimes ($43$ -- $77$\,s) (Figs.\,\ref{fig:custom2}\, -- \,\ref{fig:custom_sm}). As indicated in Sect.\,\ref{sec:disc-simulations}, a shock wave propagates through the chromosphere with typical vertical speed \hea{on} the order of $\sim\,10$\,km\,s$^{-1}$ and only a relatively small apparent motion within a few tens\,km\,s$^{-1}$ is therefore expected. Event A -- C show apparent speeds between $\sim 12$ -- $22$\,km\,s$^{-1}$ while event D appears stationary (Fig.\,\ref{fig:apparent_motion} and Figs.\,\ref{fig:custom2}\, -- \,\ref{fig:custom_sm}), which all that lies within the regime expected for shock waves. Lastly all of the locations of the events only show a very small absolute magnetic field strength (Fig.\,\ref{fig:mag_cont_mask}). The fulfillment of the criteria signatures of a shock wave by the events in the ALMA observations do suggest that their signatures could origin in shock wave events.

The evolution of the surroundings of an event can give clues on the local atmospheric structure. At the position of event~A, the brightness temperature increases and decreases repetitively independent of the surroundings (Fig.\ref{fig:custom2}), which could be interpreted as a case of a wave train with multiple shock waves being excited at the same location in the photosphere below \citep[see, e.g.,][]{2004A&A...414.1121W}.
In event~D, it is possible to see the indications of a repetitive shock wave pattern at the centre location (Fig. \ref{fig:custom_sm}h). Although, the previous $T_\mathrm{b}$ peaks are relatively weak.

The low spatial resolution makes it challenging to estimate motions on the short timescales of the lifetimes of the events. Further, the apparent motion of an event might be difficult to estimate if it is not well isolated. The low apparent motion of event D could possibly be explained by that it is unresolved. Though, the significant apparent motion spanning several pixels of the brightest points in example A -- C over their lifetimes brings reliability to the speed estimates.

The approach of looking for variations in the brightness temperature at a fixed location (time evolution for each pixel) gives in principle indications to all kinds of dynamical small-scale features. 
Out of the 552 registered events, it seems that most of the events exhibit an increase and decrease in brightness temperature at a more or less fixed (horizontal) location (see Figs.\,\ref{fig:events_coordinates} and \ref{fig:apparent_motion}), as expected for a mostly vertically propagating shock wave. The peaks in the pixels associated with an individual event occur very close in time, \hea{no more than a few seconds,} indicating that the peaks are caused by the same underlying physical event. A majority of events show a speed of apparent motion very close to zero which could probably be a result of low spatial resolution with under-resolved events and a slight shift towards higher velocities could be expected with an increase in spatial resolution.

Many of the $552$ detected events (Figs.\,\ref{fig:peak_lifetime_mag} -- \ref{fig:apparent_motion}), fulfill the aforementioned criteria which might suggest they do origin in shock waves. As a direct result of the selection process of the events, all of them show a high variation in brightness temperature. Approximately $95\%$ of the events show an absolute \hea{LOS} magnetic field strength lower than $20$\,G and an apparent speed of less than $30$\,km s$^{-1}$. Further, as discussed in Sect. \ref{sec:disc-spatial resolution}, it is very plausible that the lifetimes indicated by the 1D simulations (Sect. \ref{sec:disc-simulations}) are underestimated and that longer lifetimes would be expected in the observations. 
Assuming that lifetimes could rather be as long as 
$100$\,s would result in more than half of all detected events potentially being produced by shock waves. However, to determine to what degree the lifetimes are underestimated, further studies need to be done using more realistic numerical models.

\subsection{Numerical simulations -- 1D vs. 3D}
\label{sec:discuss_1dvs3d} 

The shape of the temporal $T_\mathrm{b}$ profile of a shock wave signature in a 3D atmosphere \hea{often} deviate rather much from the characteristic shape that is indicated in the 1D simulations. The shock wave is propagating through an inhomogenous 3D atmosphere where disturbances from previous dynamics have made an imprint or the shock wave might be interfered by other simultaneous dynamics. \hea{Not only a severely change of the amplitude of the $T_\mathrm{b}$ peak, but the temporal} profile might also be severely deformed as a result of the spatial smearing, \hea{(see Fig.\,\ref{fig:bifrost_ex})}. The temporal profile is thus very dependent on the surrounding structure but can at least be used to validate clear cases. The numerical RADYN simulations in this study provide an important insight into the expected signatures of shock waves as they could be observed in mm-wavelengths with ALMA. On the other hand, the restriction to one spatial dimension is severe in view of the complex three-dimensional structure of the solar atmosphere. The one-dimensional numerical models thus only provide a first approximation of the radiative signatures of a propagating shock wave.

\hea{A more extensive study of 2D brightness temperature} maps from numerical 3D simulations would provide a more realistic estimate and indication of the appearance of shock wave signatures at mm wavelengths as seen by ALMA. With these, the complexity of the 3D atmosphere would be accounted for. Also the degree of degradation of dynamical fine structure that comes with limited spatial resolution can be determined that, by nature, is not possible with a one-dimensional approach.


\section{Conclusions and Outlook} \label{sec:conc}

Analysing an ALMA Band~3 ($\sim3$\,mm) observation of disk-centre quiet Sun region, we detected 552 small-scale dynamic events.
An extensive study of the physical parameters of all events was carried out in order to characterise the events and see statistical trends. 
The possibility that the detected signatures comes from shock wave events is examined. For that purpose, \hea{predictions of observable shock waves signatures at mm wavelengths from} one-dimensional atmospheric models are used. 
The detected events show a large spread in brightness temperature excess ($\Delta T_\text{b}$) reaching up to $\sim 1200$\,K for some very strong events and lifetimes between $\sim 43$ -- $360$\,s. The typical values, represented by approximately half of the events, are $\sim450$ -- $750$\,K and $\sim55$ -- $125$\,s. Most of the events, however, show a $T_\mathrm{b}$ excess that is smaller than predicted by the \hea{1D} simulations.

We conclude that the restriction to one spatial dimension is a severe limitation that cannot properly account for the small-scale chromospheric pattern as a product of complex 3D wave propagation. 
Observations with the limited spatial resolution of ALMA Band~3 thus are subject to spatial smearing of the pattern, resulting in a strong reduction of the observable brightness temperature excess ($\Delta T_\text{b}$). 

\swedt{Simulations of 3D shock wave propagation   shown here and in previous studies \citep{2007A&A...471..977W,2015A&A...575A..15L,2020A&A...635A..71W}}  demonstrate that the brightness temperature variations at mm wavelengths are strongly reduced as a result of limited spatial resolution as currently achieved with ALMA in Band~3. \hea{The magnitude of the dynamic $T_\mathrm{b}$ events seen in the observations, could be considered as a lower limit.} Based on these results, it is likely that the observed parameters are in line with the values predicted by the simulations when properly accounting for the dynamic 3D structure of the solar chromosphere and the effect of limited spatial resolution during the observation. With this in consideration, it seems plausible that many of the detected events have signatures originating in propagating shock waves. A more detailed study based on \hea{3D numerical simulations is needed} and will be presented in a forthcoming paper.


\section*{Acknowledgments}
This work is supported by the SolarALMA project, which has received funding from the European Research Council (ERC) under the European Union’s Horizon 2020 research and innovation programme (grant agreement No. 682462) and by the Research Council of Norway through its Centres of Excellence scheme, project number 262622. 
This paper makes use of the following ALMA data: ADS/JAO.ALMA\#2016.1.00423.S. ALMA is a partnership of ESO (representing its member states), NSF (USA) and NINS (Japan), together with NRC(Canada), MOST and ASIAA (Taiwan), and KASI (Republic of Korea), in co-operation with the Republic of Chile. The Joint ALMA Observatory is operated by ESO, AUI/NRAO and NAOJ. We are grateful to the many colleagues who contributed to developing the solar observing modes for ALMA and for support from the ALMA Regional Centres. We also acknowledge collaboration with the Solar Simulations for the Atacama Large Millimeter Observatory Network (SSALMON, http://www.ssalmon.uio.no).

\bibliographystyle{aa}
\bibliography{henk.bib}

\appendix
\section{\hea{Example of shock wave in Bifrost 3D simulation}}
\label{sec:appendix:bifrost}
\begin{figure*}[tp!]
\sidecaption
\includegraphics[width=12cm]{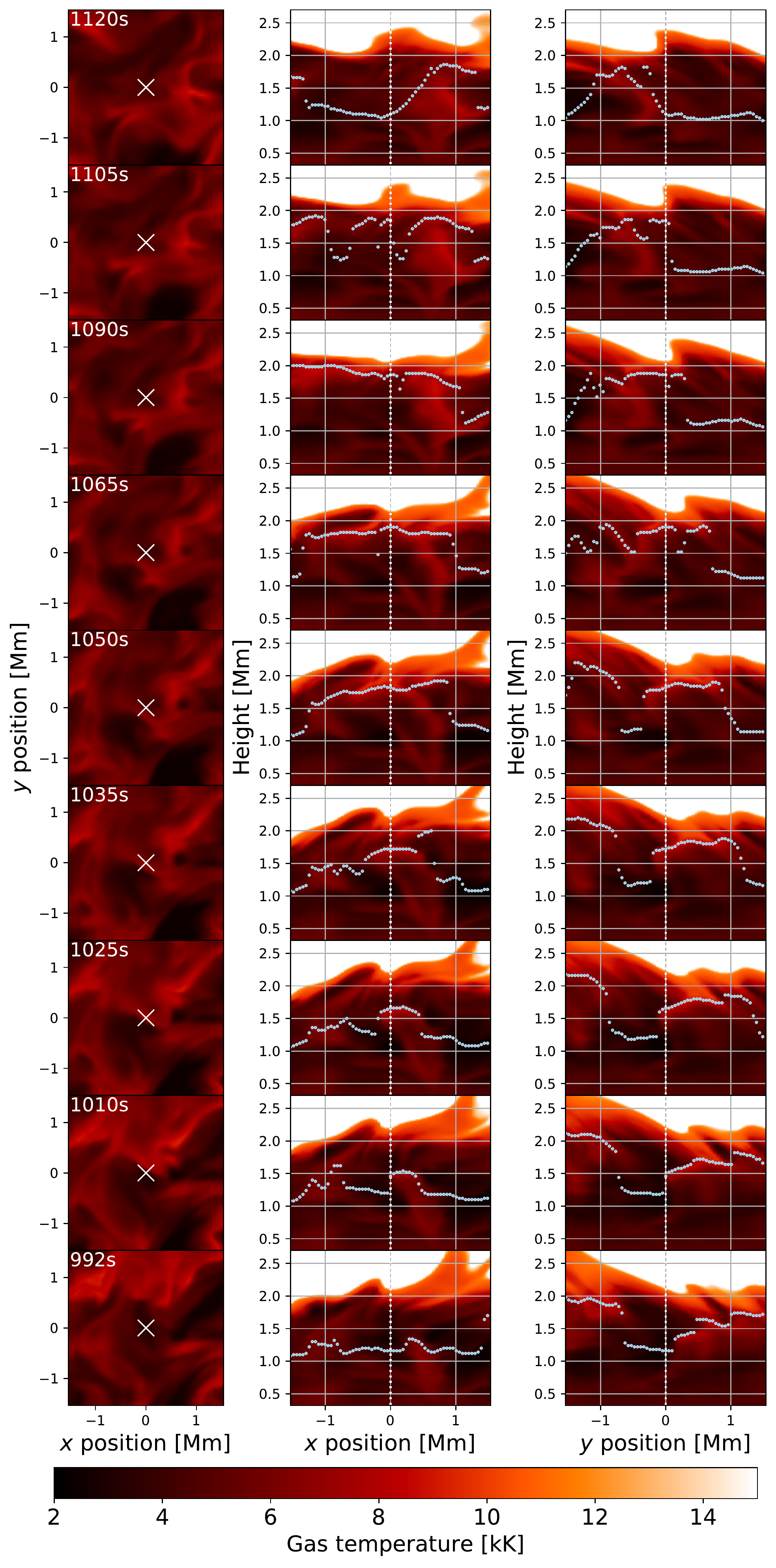}
\caption{\hea{Illustration of propagating shock wave in the Bifrost 3D simulation. Gas temperature surrounding the shock wave example for 9 time steps between $t=992$\,s and $t=1120$\,s of the simulation run. The leftmost column shows horizontal cuts (top-view) at a height of $z=1.5$\,Mm, the middle and rightmost columns vertical cuts with the height versus the $x$- and $y$-coordinates, respectively. The blue markers show the formation height of unity optical depth $\tau=1.0$ for the mm-wavelenghts of ALMA band 3 ($2.78-3.26$\,mm). The white crosses and the vertical white dashed line at $(x,y)=(0,0)$ marks the location of the sampled $T_\mathrm{b}$ signature in Fig.\,\ref{fig:bifrost_ex}c. At $t=992$\,s the shock front is visible and the mm-wavelength intensities at $\tau=1.0$ is coupled with it at $(x,y,z)=(0.3,0.3,1.2)$\,Mm. The front is propagating mostly upwards and inwards towards $(x,y,z)=(0,0,1.6)$\,Mm at $t=1025$\,s. The $T_\mathrm{b}$ signature is peaking in magnitude at this location around $t=1050$\,s, when a cooler post-shock region is visible where the front has passed. The mm-wavelength intensities are tracking the shock front upwards to $z\approx2.0$\,Mm around $t=1090$\,s and thereafter decouples and sampling the post-shock region ($t=1120$\,s). Compare with \cite{2020arXiv200805324E} for more detailed analysis of propagating shock waves in the Bifrost simulation.}}
\label{fig:appendix:bifrost}
\end{figure*}

\swedt{In Fig.~\ref{fig:appendix:bifrost}, the evolution of the gas temperature in a small region of the solar atmosphere as part of a 3D magnetohydrodynamical simulation with the Bifrost code  \citep{2011A&A...531A.154G,2016A&A...585A...4C} is shown. 
The displayed data exhibit  an example of shock wave propagation. 
This example is used to illustrate the connection between the atmospheric thermal structure and the observable brightness temperature in mm-wavelength radiation maps as 
 discussed in Sect.\,\ref{sec:disc-3D-simulations}.}

\section{\hea{Detailed study of additional examples of brightness temperature events in the ALMA data}}
\label{sec:appendix:events}

\hea{An additional couple of examples of $T_\text{b}$ events found in the observational data are shown in detail. These events are marked as 'B' and 'D' in the figures in the main text.}

\begin{figure*}[tp!]
\includegraphics[width=\textwidth]{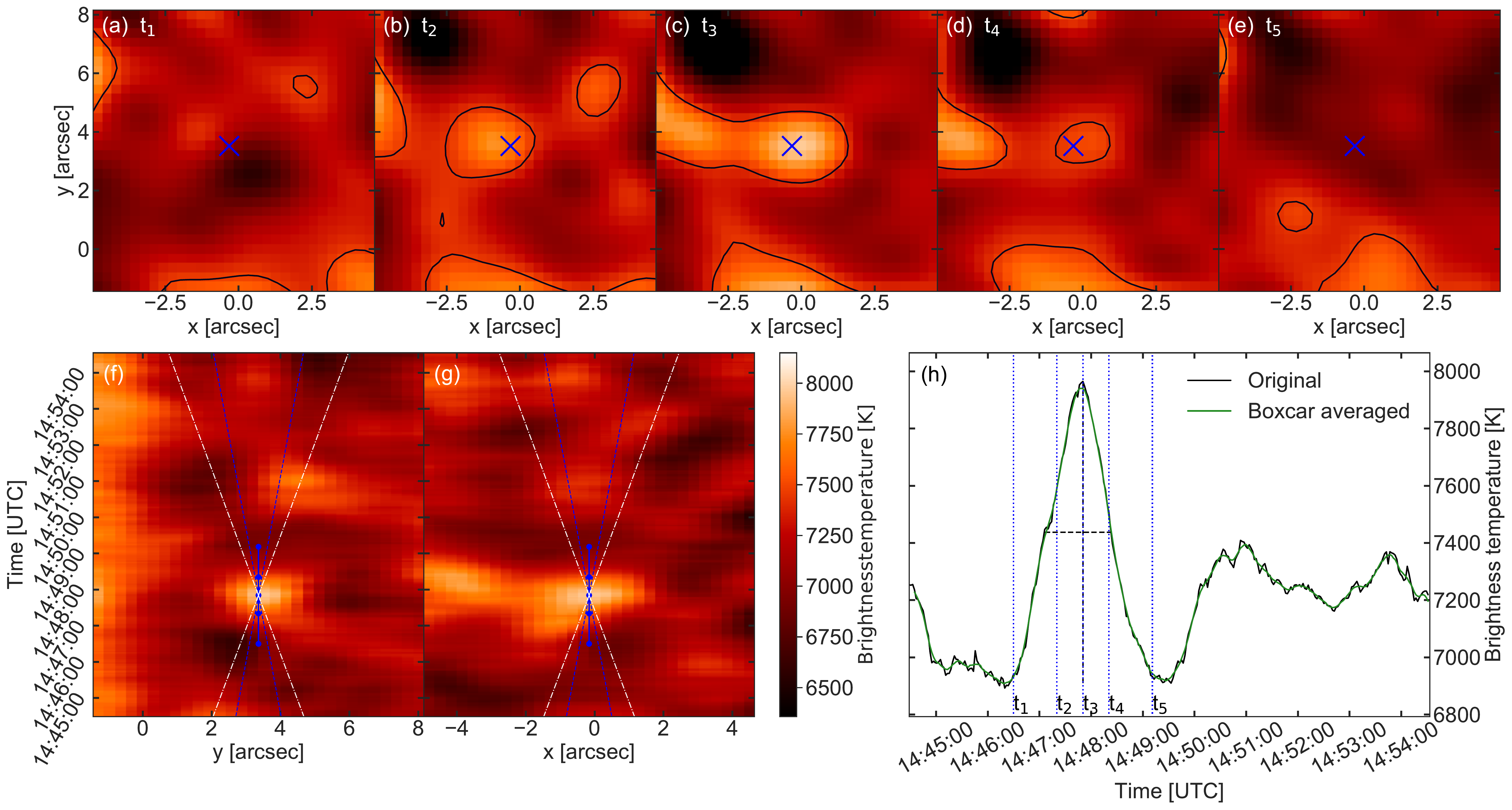}
\caption{Detailed study of event B.  
\textbf{a-e)} Close ups of the surroundings brightness temperatures at different time steps, $t_1$ to $t_5$ from left to right, through the shock wave event. 
The time of the $T_\mathrm{b}$ peak is marked by $t_3=14$:$47$:$50$. $t_1, t_2, t_4$ and $t_5$ mark $-80,-30,30$ and $80$\,s from the peak, respectively.
The contour lines marks the half maximum of the maximum $\Delta T_\text{b}$ peak and the blue crosses mark the center location.
\textbf{f-g)} Space-time diagrams for a vertical and horizontal slit across the FOV at the center location. The centre location for time steps $t_1$ to $t_5$ is marked with blue dots connected with a line. Velocity slopes for $10$ and $20$\,km\,s\,$^{-1}$ are indicated by blue dotted and white dashed lines, respectively.
\textbf{h)} The time evolution of the brightness temperature of the center location, where the time steps $t_1$ to $t_5$ are indicated by blue dotted lines. Both the original data (black) and the averaged data (green) are shown. The horizontal and vertical dashed lines marks the event lifetime and brightness temperature excess.}
\label{fig:custom1}
\end{figure*}

Event~B (see Fig.\,\ref{fig:custom1}), is located a bit closer to the magnetically stronger areas in the upper FOV at $(x,y)=(-0.2\arcsec,3.4\arcsec)$ right on the border of the network mask, though the absolute magnetic field strength is low at about $56$\,G. Event~B shows a similar evolution in brightness temperature as event A. There is a rise in $T_\mathrm{b}$ from about $6890$\,K to $7960$\,K in $90$\,s followed by a decrease to about $6910$\,K in the course of $100$\,s. Event~B has thus a brightness temperature excess of $\Delta T_\text{b} = 1055$\,K and a lifetime of $77$\,s. The profile of the temporal evolution of the brightness temperature (Fig.\ref{fig:custom1}h) shows a very similar shape as the fiducial run of the one-dimensional simulations. Event B, roughly with a size of $\sim\,3\arcsec$ at the temperature of FWHM of the main peak, shows an apparent motion of approximately $12$\,km\,s$^{-1}$.

\begin{figure*}[tp!]
\includegraphics[width=\textwidth]{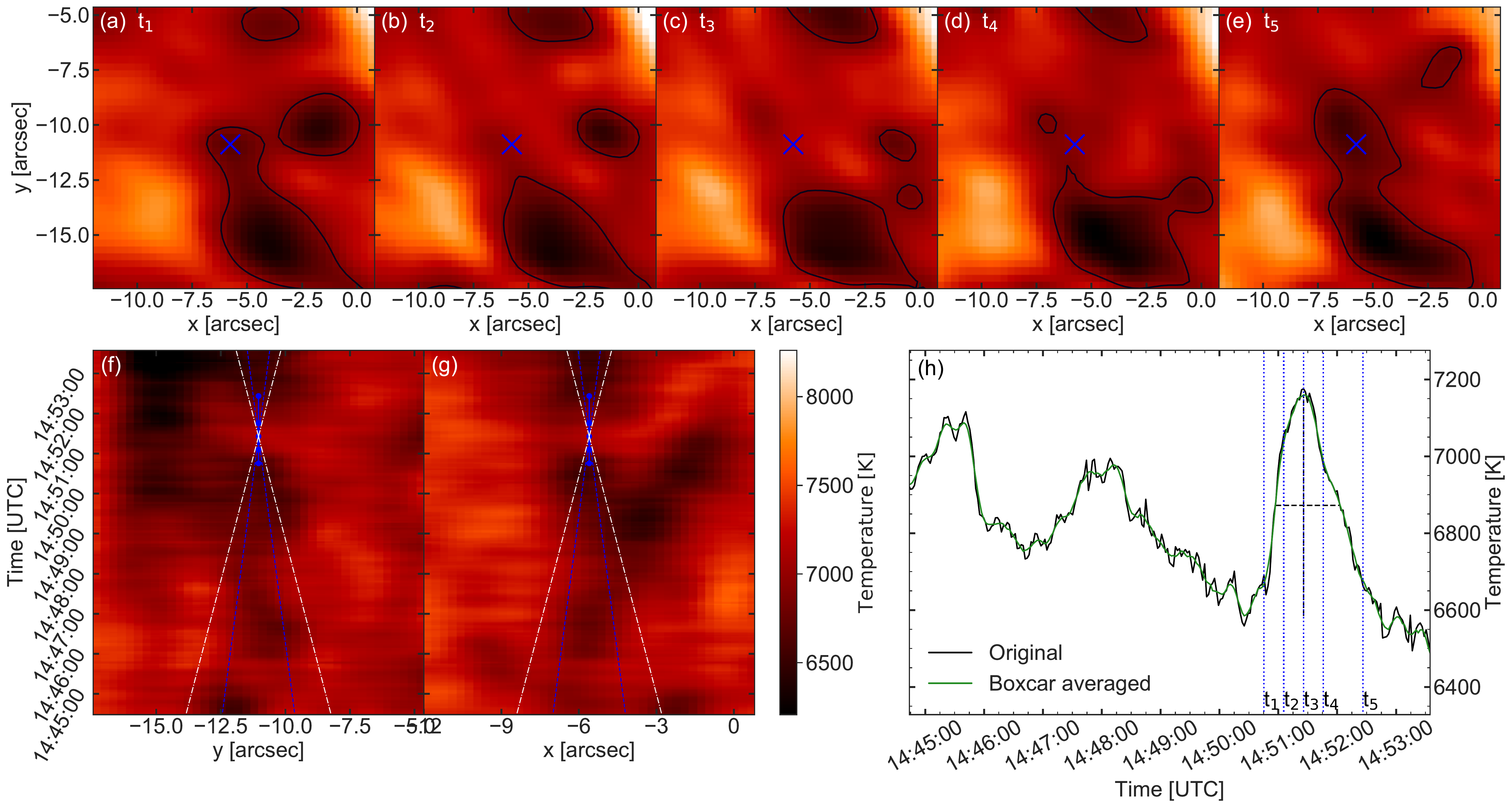}
\caption{Detailed study of event D. Description as for event B in Fig.\,\ref{fig:custom1}. The time of the $T_\mathrm{b}$ peak is marked by $t_3=14$:$51$:$25$. $t_1, t_2, t_4$ and $t_5$ mark $-40,-20,20$ and $40$\,s from the peak, respectively.}
\label{fig:custom_sm}
\end{figure*}

Event~D (Fig.\,\ref{fig:custom_sm}) at a location of $(x,y)=(-11.0\arcsec, -5.6\arcsec)$ shows a rapid increase in brightness temperature from $6585$\,K to $7160$\,K followed by a decrease  to $6450$\,K. The resulting amplitude and lifetime are $\Delta T_\text{b} = 575$\,K and $65$\,s, respectively. The event appears horizontally stationary without any strong indications of apparent motion. At the specific location there are hints of previous peaks of brightness temperature excess. Weak signatures, merely $200$ -- $300$\,K as seen in Fig.\,\ref{fig:custom_sm}h approximately 6\,min and 3.5\,min before event D.

\end{document}